\documentclass[letterpaper,twocolumn,10pt]{article}
\usepackage{usenix2019_v3, epsfig}

\usepackage{booktabs} 
\usepackage{epstopdf}
\usepackage{adjustbox}
\let\labelindent\relax
\usepackage{enumitem}
\usepackage{multirow}
\usepackage{bm}
\usepackage{float}
\usepackage[ruled,linesnumbered]{algorithm2e}
\usepackage{graphicx}
\usepackage{pythonhighlight}
\usepackage[flushmargin]{footmisc}
\usepackage{authblk}
\newcommand{\eg}{e.g.,}
\newcommand{\ie}{i.e.,}
\newcommand{\etal}{\textit{et al.}}
\usepackage[T1]{fontenc}
\usepackage{listings}
\usepackage{color}
\usepackage{xcolor}
\usepackage{hyperref}
\usepackage{placeins}
\usepackage{url}

\usepackage{comment}
\usepackage{xspace}

\renewcommand{\paragraph}[1]{\vspace{2pt}\noindent\textbf{#1}}

\definecolor{mGreen}{rgb}{0,0.6,0}
\definecolor{mGray}{rgb}{0.5,0.5,0.5}
\definecolor{mPurple}{rgb}{0.58,0,0.82}
\definecolor{backgroundColour}{rgb}{0.95,0.95,0.92}

\lstdefinestyle{CStyle}{
    commentstyle=\color{mGreen},
    keywordstyle=\color{magenta},
    stringstyle=\color{mPurple},
    basicstyle=\scriptsize\ttfamily,
    breaklines=true,
    captionpos=b,
    keepspaces=true,
    otherkeywords={status_t, ArrayObject,uint32_t},
    numbers=left,
    numbersep=1pt,
    showspaces=false,
    showstringspaces=false,
    showtabs=false,
    tabsize=2,
    language=C
}

\newcommand{\gl}[1]{\textcolor{red}{#1}}
\newcommand{\condr}{{conditional register}\xspace}
\newcommand{\condrs}{{conditional registers}\xspace}
\newcommand{\kb}{{\textit{KB}}\xspace}
\newcommand{\sys}{{\textit{$\mu$Emu}}\xspace}
\newcommand{\ppim}{{P$^{2}$IM}\xspace}
\newcommand{\hal}{{HALucinator}\xspace}

\newcommand{\cm}{Cortex-M\xspace}
\newcommand{\twrite}{T0\xspace}
\newcommand{\tone}{T1\xspace}
\newcommand{\ttwo}{T2\xspace}
\newcommand{\tthree}{T3\xspace}

\usepackage{tikz}
\usepackage{amsmath}

\begin{document}

\title{Automatic Firmware Emulation through Invalidity-guided Knowledge
Inference (Extended Version)}

\author[1]{Wei Zhou}
\author[2]{Le Guan}
\author[3]{Peng Liu}
\author[ ]{Yuqing Zhang$^{1,4,5}$}
\affil[1]{\textit{National Computer Network Intrusion Protection Center, University of Chinese Academy of Sciences, China}}
\affil[2]{\textit{Department of Computer Science, University of Georgia, USA}}
\affil[3]{\textit{College of Information Sciences and Technology, The Pennsylvania State University, USA}}
\affil[4]{\textit{School of Cyber Engineering, Xidian University, China}}
\affil[5]{\textit{School of Computer Science and Cyberspace Security, Hainan University, China}}
\renewcommand*{\Affilfont}{\normalsize\it} 
\renewcommand\Authands{ and }

\maketitle
\pagestyle{empty}  
\thispagestyle{empty} 

\begin{abstract}
Emulating firmware for microcontrollers
is challenging due to the tight coupling between the hardware and firmware.
This has greatly impeded the application of
dynamic analysis tools to firmware analysis.
The state-of-the-art work 
automatically models unknown peripherals 
by observing their access patterns,
and then leverages heuristics to calculate the appropriate responses
when unknown peripheral registers are accessed.
However, we empirically found that this approach and the corresponding heuristics are frequently insufficient to emulate firmware.
In this work, we propose a new approach called \sys to emulate firmware with
unknown peripherals.
Unlike existing work that attempts to build a general model
for each peripheral, our approach learns how to correctly emulate firmware execution at individual peripheral access points.
It takes the image as input and symbolically 
executes it by representing unknown peripheral registers as symbols.
During symbolic execution, it
infers the rules to respond to unknown peripheral accesses.
These rules are stored in a knowledge base, which is referred to during the
dynamic firmware analysis.
\sys achieved a passing rate of 95\%
in a set of unit tests for peripheral drivers without any manual assistance.
We also evaluated \sys with real-world firmware samples and new bugs were discovered.






\end{abstract}


\vspace*{-2mm} 
\section{Introduction}
\vspace*{-2mm} 

The rapid emergence of Internet of Things (IoT) technology makes microcontrollers (MCUs) an increasingly serious security concern. 
Since most real-world IoT devices run on MCU-based SoCs (System on Chip)
and since MCUs lack many security threat mitigation mechanisms available on PC/mobile platforms, 
many recent security incidents have been related to MCU security.
In MCU firmware, the main task
runs in an infinite loop that constantly monitors and handles external events.
The task code implements the core logic of the application and integrates necessary libraries,
such as the TCP/IP stack and MQTT protocol.
The external events on the other hand,
are abstracted by the kernel (if any) and peripheral drivers.
The mentioned security incidents were the result of vulnerabilities within either
the task code~\cite{bugs_of_tcpip,urgent11} or the driver code~\cite{DSPIdriver, garbelini2020sweyntooth, ruge2020frankenstein}.





Dynamically analyzing the task code in MCU firmware is challenging,
since its execution depends on (1) the runtime environment constructed during
device bootstrapping, and (2) the driver functions directly invoked by the task.
For example, to find a bug in the task code caused by improper handling of input from the UART interface,
the driver code of the UART peripheral should be executed without hanging or crashing the firmware.
To satisfy these requirements,
an emulator must emulate the logic of diverse peripherals 
on real-world MCUs. 
For example, when the firmware reads a register of a custom-made peripheral,
the emulator should return an appropriate value depending on the current peripheral status. 
Given the high-diversity in the ecosystem of MCU SoCs in the market, it
would require a huge amount of manual effort to develop an emulator for 
(multiple types of MCU SoCs in) the ecosystem, if  
the logic of diverse peripherals could not be automatically handled.




To address this challenge, three lines of research are being conducted.
First, 
several solutions~\cite{cadar2008klee,koscher2015surrogates,talebi2018charm,muench2018avatar2} propose to forward
the interactions with unsupported peripherals to the real hardware.
However, these hardware-in-the-loop approaches cannot be used for large-scale automatic dynamic analysis. 
Second, abstraction-based approaches 
side-step the problem of peripheral emulation by
leveraging the abstraction layer available on firmware.
For example, by emulating such an abstraction layer
in Linux kernel, 
many Linux-based firmware binaries can 
be emulated~\cite{shoshitaishvili2015firmalice, chen2016towards, costin2016automated, kim2020firmae}.
Recently, \hal~\cite{clements2020halucinator} has been proposed to automatically match
the \textit{Hardware Abstraction Layer} (HAL) APIs in firmware and replace them with host implementations.
However, 
this approach requires ecosystem-wide standardization and is 
problematic for firmware on custom-made SoCs~\cite{SmartThings, Philips, TP}.
In real-world firmware development, developers can invoke driver functions in arbitrary ways. 
It is therefore difficult to decouple the security testing of task code of firmware
from driver code execution. 
Moreover, since this approach completely skips the peripheral logic in firmware, dynamic analysis
cannot find any bugs in the peripheral drivers.
{\bf Third}, full-system emulation~\cite{fengp2020p2im, gustafson2019toward, cao2020device}
aims to emulate the entire firmware without relying on real hardware.
For example, \ppim~\cite{fengp2020p2im}, a representative approach in this  direction,
observes the access pattern of an unknown peripheral and infers its interaction model~\cite{fengp2020p2im}.
Then \ppim combines expert-provided heuristics and such 
interaction models to figure out how to infer 
the appropriate responses from peripherals.
Laelaps~\cite{cao2020device} uses symbolic execution to explore possible branches,
and then leverages heuristics to predict a ``good'' one to follow.


 


Although the third research direction has shown exciting potential for 
achieving device-agnostic emulation with high fidelity, 
based on our empirical studies, 
we still found they frequently fail to properly execute complex samples.
For example, \ppim has to blindly guess the appropriate responses
for read operations to the status registers of peripherals, which is impractical considering the large search space.
Restricted by the exploration depth,
Laelaps~\cite{cao2020device} can only find a good branch for a short period of future execution.
But this decision might not be the best in the long run.
Both of them may crash or hang the emulation.

These failures are caused by
a largely-ignored fact is that firmware emulation is 
{\bf collectively} affected by multiple peripheral
registers.  By ``collectively'', we mean that in many cases 
how one peripheral access should be handled at time $t$ is 
{\bf dependent upon} the time $t$ values of several other 
registers.
For example, in following code snippet extracted from the Ethernet driver,
the \texttt{CR} and \texttt{SR} registers
are both used to decide a branch target.
When the \texttt{SR} register was accessed, the response to it is dependent upon the value of the \texttt{CR} register at that moment.

\begin{footnotesize}
\vspace*{-2mm}
\begin{verbatim}
if (EMAC->CR & EMAC->SR == 0x1E7FF)
    Enable_Ethernet_Interrupt();
\end{verbatim}
\vspace*{-2mm}
\end{footnotesize}
Based on this key insight, the emulator should recognize how 
multiple peripheral registers can affect firmware execution and 
correspondingly decide the {\em coordinated} responses.

Meeting this requirement is challenging due to the lack of firmware semantics.
For example, \ppim observes the interaction patterns of each peripheral and
handles each peripheral access individually without taking the above-mentioned
dependency into consideration. However, the observed execution trace does not
provide enough contextual information to properly categorize registers or
calculate a coordinated response.

\paragraph{Our idea.} 
As mentioned before, to dynamically analyze task code of firmware,
it is important to emulate the hardware behaviors entirely, including those of peripherals.
Only in this way can we reach to (buggy) task code responsible for handling input retrieved from the I/O interface.
To learn peripheral behaviors and correspondingly emulate driver code,
we observe that analyzing the interaction patterns of every peripheral is actually unnecessary.
\emph{As long as we can
decide an appropriate dependency-aware 
response at each peripheral access 
point, the emulation may succeed.}  
To realize this idea, two questions need to be 
answered. How to judge whether a peripheral input is
appropriate or not? How to obtain such an appropriate
peripheral input?  In this work, we answer these questions with two
observations and correspondingly developed a system called \sys.
{\bf Observation 1:} If a response is incorrectly fed to the firmware,
the error will eventually be reflected in the execution state. In particular,
the emulation would enter an invalid state. 
{\bf Observation 2:} An invalid execution state is directly 
reflected on an invalid path. To avoid executing invalid 
 paths, we can 
 represent all the peripheral responses as symbols, and then use symbolic execution to {\bf collectively} reason about peripheral responses that can avoid such states/paths.
Through collective reasoning, we can achieve 
{\bf dependency-aware} peripheral access handling. 
Through symbolic execution, we can achieve 
{\bf constraint-satisfaction-based} response finding.




Following these two observations, we propose \sys,
a dynamic analysis tool for find bugs in the task code of firmware for ARM MCUs.
The core component of \sys is a device-agnostic emulator
aiming at emulating driver code of unknown peripherals.
We infer necessary knowledge for properly emulating a
specific firmware image using invalidity-guided symbolic execution.
Our system is comprised of two phases,
the knowledge extraction phase and the dynamic analysis phase.
In the knowledge extraction phase,
it takes the firmware-under-test as input and
mixes concrete and symbolic execution (\ie~concolic execution)
to extract essential information for the subsequent dynamic analysis phase,
The information is stored in a \textit{knowledge base} (\kb) for later references.
Replacing concrete execution with concolic execution, the proposed approach can reach deep paths and extract additional knowledge. 
Using a symbolic constraint solver also enables the proposed approach to 
accurately find the appropriate peripheral readings.
In the firmware dynamic analysis phase,
\sys matches the entries in the extracted knowledge base and responds with appropriate values
when a register of a (custom-made) peripheral is read.
The knowledge base guides the execution to always stay in valid states,
while value mutations of data registers, which can be controlled when
the attacker has access to the I/O interface, help find new execution paths
and firmware defects.




During knowledge extraction, \sys only switches to another path when the
current path is found invalid. Therefore, the path explosion problem faced by many symbolic-execution-based approaches, including Laelaps~\cite{cao2020device}, is alleviated naturally.
Moreover, knowledge (e.g., a concrete value for a particular register) extracted at an earlier time point -- if found useful -- can always be used at a later time point. This
avoids potential symbolic execution. As a result,
path explosion is further reduced and time-consuming solver invocations are minimized.
In contrast, Laelaps needs to enter expensive symbolic execution every time a peripheral register is accessed.

A notable feature of the proposed approach is that the knowledge
base built with the restricted exploration space (i.e., if the current path 
remains valid, \sys will stick to it)  
in the knowledge extraction phase can be used to 
emulate multiple valid paths  
in the dynamic analysis phase. 
This is because \sys adopts a tiered caching mechanism,
in which a cache entry uses progressively more context information 
to decide a response. Accordingly, (a) the bottom tier 
knowledge enables the emulator to use the last written value 
as the response to a peripheral read access; such values
are dynamically determined and can fork new branches. 
(b) the upper tiers use more restrictive matching rules
and therefore can record multiple branch matching rules
based on different contexts during knowledge extraction.
In the dynamic analysis phase, new paths can be emulated when contexts
are changed.

We evaluated \sys with 66 unit tests for testing the basic function of individual peripherals.
Compared with the passing rate of 79\% achieved by \ppim, \sys achieves
95\%
without any manual assistance.
With very little manual assistance,
all unit tests can be passed.
We also evaluated \sys with 21 real-world firmware samples.
Evaluation results show that \sys
is capable of emulating real-world firmware.
By bridging it with AFL, a state-of-the-art fuzzing tool,
\sys also helped us find previously-unknown bugs in the task code of
the tested samples.

In summary, we made the following contributions.
\vspace*{-1mm}
\begin{itemize}
\vspace*{-1mm}
\item We proposed using symbolic execution to emulate
  MCU firmware without relying on real hardware.
  We achieved this through an invalidity-guided recursive knowledge extraction algorithm.
  The cached results in turn allow us to build a knowledge base
  for the firmware used for dynamic analysis.
  \vspace*{-1.5mm}
  \item We implemented our idea on top of S2E. We 
  show the practicality of our approach by 
  evaluating it on a collection of 21 real-world firmware samples covering more than 30 different kinds of peripherals with several popular MCUs.
 \vspace*{-1.5mm}
  \item
  We also integrated a modified AFL fuzzer with \sys. Through fuzzing analysis, we reproduced existing bugs as well as found new bugs.
  \sys is open source at \url{https://github.com/MCUSec/uEmu}.
\vspace*{-1.5mm}

  
  

\end{itemize}
\vspace*{-2mm}
\vspace*{-2mm}
\section{Background}
\vspace*{-1mm}

\subsection{MCU Peripherals}

MCUs have widely adopted in power-effective embedded devices such as drones,
robots and programmable logic controllers (PLCs). 
Their firmware typically comprises the task code (including the core logic implementation and dependent libraries),
the kernel code (if any), and the driver code for peripherals.
MCU peripherals are mainly used to communicate with the external world.
There are
two types peripherals, on-chip peripherals and off-chip peripherals. The
functions of on-chip peripherals are invoked by writing to or reading from
peripheral registers, which are typically memory-mapped into the system
memory. For example, on ARM Cortex-M MCUs, peripheral registers are mapped
from \texttt{0x40000000} to \texttt{0x5fffffff}. The values of peripheral
registers change non-deterministically depending on the internal logic of the
peripheral. To increase efficiency, using interrupts is a common practice.
Off-chip peripherals are oblivious to the MCU core. They are connected to the
MCU core via on-chip peripherals, which  serve as proxies between the firmware
and off-chip peripherals. For example, the SPI peripheral, which is a
general-purpose communication bus, is commonly used to connect EEPROM and
BlueTooth peripherals.

MCU peripherals are very diverse. On the one hand, there are hundreds of
different types of peripherals dedicated for different tasks. On the other
hand, even for the same type of peripheral such as UART, manufacturers often
implement it in customized ways. This diversity imposes a major obstacle for us
to emulate a previously-unseen firmware image. Specifically, the internal
logic of each peripheral has to be accurately and individually emulated.

\vspace*{-3mm}
\subsection{Dynamic Symbolic Execution and S2E} 
\vspace*{-2mm}
\label{sec:s2e}

Symbolic execution~\cite{king1976symbolic} is a powerful automated software testing and analysis technique. 
It treats program
inputs as symbolic variables and simulates program execution so that all variables are represented as symbolic expressions.
Dynamic symbolic execution (a.k.a. concolic execution)
combines concrete execution and symbolic execution and inherits
the advantages of both. It has been widely used to finding deep vulnerabilities in commercial
software~\cite{cadar2008klee, godefroid2008automated}.



S2E~\cite{chipounov2011s2e} is one of the most popular open-source symbolic execution platforms. Since it is based on QEMU, 
it enables full system symbolic execution and thus supports
testing both user-space applications as well as drivers.
More importantly, S2E exposes useful APIs to extend its functionality.
An active community constantly writes and 
maintains many useful S2E plugins for performance improvement (\eg~better
state pruning algorithms) or new program analysis tool development.
Although QEMU supports multiple architectures,
the latest S2E only supports emulating x86/x86-64 architecture~\cite{armsupport}.
In the following paragraphs, we introduce necessary
technical background for understanding this paper.

\paragraph{CPU Emulation and Hardware Emulation.}
The original S2E is tightly coupled with QEMU.
It leverages the Dynamic Binary Translation (DBT) of QEMU to emulate
CPU and combines it with KLEE~\cite{cadar2008klee} for concolic execution.
The hardware such as peripherals is emulated by QEMU.


\paragraph{KVM Interface.}
S2E developers found it tedious to update with
the upstream QEMU. Since version 2.0, they 
reconstructed the S2E architecture to de-couple it from QEMU using the KVM interface.
The new S2E only uses QEMU as a KVM client for hardware emulation,
and maintains the concolic execution engine by its own (in essence, the old DBT code in QEMU).
The concolic execution engine exposes a KVM interface for the QEMU hardware
emulator to invoke.
As a result, as long as the KVM interface is stable, when QEMU is updated,
S2E  can also be easily updated to benefit from the ever-improving emulation capability of QEMU.

\paragraph{Effective Concolic Execution.}
S2E extracts CPU emulation and DBT functions from the original QEMU and extends them with KLEE for concolic execution.
It can automatically switch between the symbolic execution engine and concrete
execution engine.
Specifically,
when a memory location containing symbolic data is de-referenced, S2E re-translates
the current translation block into LLVM IR and switches to KLEE.
When there is no longer any symbolic data in any registers, it will switch back to the DBT engine.
When encountering a branch whose target is determined by a symbol, S2E forks
a new execution state. 
S2E explores each execution state independently.
To achieve this goal, S2E maintains dedicated memory to store the hardware state for each state.

\vspace*{-4mm}
\subsection{Terminology}
\label{sec:terms}
\vspace*{-2mm}
\paragraph{Branch.}
A branch instruction is the last instruction in a basic block.
It causes the program to deviate from its default behavior of executing instructions in order.

\paragraph{Branch Target.} 
Depending on whether a branch is taken or not,
there are typically two branch targets to be executed following
the branch instruction.
In this paper, we mainly consider conditional branches in which
one or more peripheral readings decide which branch target
to follow.

\paragraph{Conditional Registers.} At each branch, one or more peripheral
registers decide the branch target. We call these registers as \condrs.

\paragraph{Execution Path/Trace.}
An execution path/trace refers to 
a dynamic flow in the control-flow graph of the program.
It starts from the program entry point and ends at an exit point.
In a firmware image, two different execution paths/traces
are created when the execution faces a branch which is determined by a peripheral reading.
In this paper, we use path and trace interchangeably
to refer to the dynamic control flow of
the firmware.


\paragraph{Execution State.}
An execution state is a break point in an execution path.
It contains a program's memory, registers, peripheral
states, etc.
S2E switches among execution states to explore
the program. When the firmware exits, the current execution
state corresponds to a unique execution path.

\paragraph{Invalid Execution State.}
An invalid execution state disrupts normal firmware execution,
including crashing or stalling firmware execution,
and skipping designed operations.
At the core of our system
is an exploration algorithm that constantly detects and avoids invalid execution states
caused by wrong peripheral readings.


\paragraph{Valid Execution State.}
Valid execution states are execution states that are not invalid.
By responding to the firmware execution with the values stored
in the knowledge base, \sys keeps the firmware emulation in valid execution states.




\vspace*{-3mm}
\section{Overview}
\vspace*{-2mm}

\label{sec:overview}

The goal of \sys is to find bugs in task code of firmware related to improper handling
of malformed input retrieved from data registers of the I/O interfaces.
Therefore, it needs to emulate the peripheral
drivers, especially those related to I/O, by
automatically generating \emph{appropriate} responses
when an unknown peripheral register is accessed.
However, we cannot guarantee the same readings
as real peripherals.
Rather, the provided (response) values should 
pass the firmware’s internal checks so that the firmware execution could reach 
a useful state for practical security analysis. 

\vspace*{-3mm}
\subsection{High-level Idea}
\vspace*{-1mm}

Our work is based on three insights.
First, in MCU firmware, \condr readings often directly influence
the execution path.
Second, by representing the peripheral registers as symbols,
the relationship between the peripheral register and the path
can be captured by symbolic expressions.
Third, if an incorrect path is selected, the firmware will
reach an invalid state.
Therefore, our approach represents all the readings from unknown 
peripherals as symbols,
and leverages symbolic execution and 
an invalid state detection mechanism to automatically
extracts knowledge about how to respond to peripheral accesses.
The extracted information includes
(1) a knowledge base regarding how to respond to unknown peripheral accesses 
so that the execution will stay valid;
and (2) a set of identified data registers used for I/O operations.

The knowledge base is a cache of knowledge learned from symbolic exploration.
In a firmware execution, the same peripheral register could be 
accessed many times and the peripheral returns a 
value depending on the current hardware state machine.
In \sys, we model an approximate hardware state machine using 
peripheral context (\eg~the current function arguments),
and use this context to match a cache entry.
Specifically, in the knowledge extraction phase  
\sys starts with a simple matching rule aiming to match many 
similar peripheral accesses.
However, when the cached value is proven wrong (by invalidity checks in future execution), it is rejected and upgraded.
The upgraded matching rule considers complex execution context and 
thus only matches specific peripheral accesses with the same context.
In short, a cache entry uses progressively more context information 
to decide a response.
While the simple matching rule helps \sys quickly reduce the exploration space of symbolic execution, the context-aware matching rule kicks in when the simple one cannot handle the complex situations.

\vspace*{-2.5mm}
\subsection{Threat Model} 
\vspace*{-1.5mm}

\sys is a bug-driven firmware emulator. The ultimate goal is to find software
bugs in the task code of firmware that can be leveraged to hijack the control flow of the firmware, steal
confidential information, launch DoS attacks, etc. In this paper, we focus on
finding memory-related bugs by fuzzing. However, the capability of emulating
firmware execution allows \sys to be used with other dynamic analysis tools.
The attacker is assumed to  have access to standard I/O interfaces of the
device, \eg~the SPI or UART, and thus can feed malformed data to these
interfaces. We do not consider powerful attackers who can cause circuit-level
manipulation, including arbitrarily changing the values of control registers
or status registers. Therefore, \sys calculates appropriate values for accesses
to control/status registers so that peripheral drivers avoid entering error
handling states. It also identifies data registers used in I/O, which can be
controlled by the attacker. During dynamic analysis, we consider the input to
the data registers as untrusted and find memory corruptions caused by the
malformed input.

\begin{figure*}[t]
\centering
\includegraphics[width=.9\textwidth]{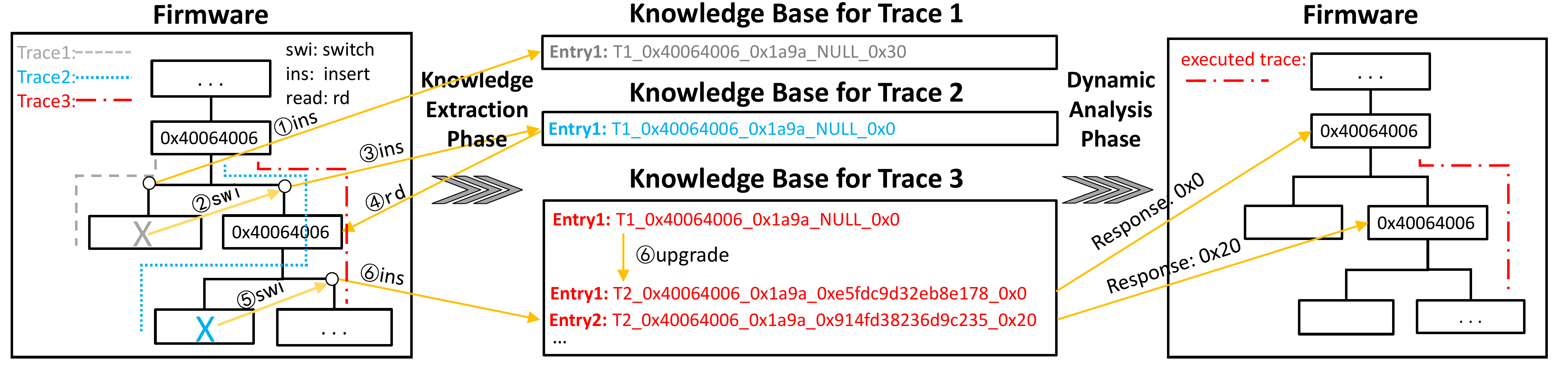}
\vspace*{-5mm}
\caption{A Running Example of \sys.}
\label{fig:overview}
\flushleft
\scriptsize{
For easy representation, only caching rule \tone and \ttwo are mentioned.
In the knowledge base, each entry includes the following
information: 1: the caching rule; 2: the address of involved register;
3: the PC at which the register is accessed; 4: the hash of the context information;
and 5: the cached value.
}
\vspace*{-3mm}
\end{figure*}

\vspace*{-2.5mm}
\subsection{Our Approach} 
\vspace*{-1.5mm}



\sys is a two-phase system for emulating and analyzing
MCU firmware (Figure~\ref{fig:overview}). 
For each firmware image, we first run a knowledge extraction phase in which
a knowledge base regarding how to respond to peripheral accesses is built.
Moreover, a set of data registers used for I/O operations are identified.
In the second phase, we use dynamic analysis approaches to test the firmware. When
a custom-made peripheral is accessed, the appropriate response value is directly obtained by 
referring to the \kb. 
Accesses to data registers are directly bridged
to the analysis tools such as a fuzzer to test the task code of the firmware.
If a query does not match any cache entries in the \kb, the knowledge extraction phase needs to be incrementally re-executed to enrich the \kb.

\paragraph{Knowledge Extraction Phase.}
At the core of the knowledge extraction phase is an \emph{invalidity-guided} symbolic execution engine.
During symbolic execution,
peripheral readings calculated (via a constraint solver) during previous exploration
are cached in \kb using a \emph{tiered caching strategy}.
When a register of an unknown peripheral is accessed,
\sys represents it as a symbol.
If this symbol directly impacts a branch target during symbolic execution,
\sys chooses a default branch target
and caches
the solved values for later accesses.
The cached values help the symbolic execution engine decide a favorable branch
target when the same peripheral is accessed later on.
Specifically, the cached value is used in a tentative concrete computation
to decide the corresponding branch target.
We adopt a tiered caching strategy.
\sys starts
with a simple matching rule aiming to 
let a cache entry match as many similar peripheral accesses as possible.  
If later we find the cached value was wrong,
we reject it and upgrade the matching rule for the corresponding
peripheral register.
The indicator for a wrong cache entry
is that the execution state becomes invalid (Section~\ref{sec:rejection}).
The upgraded matching rule captures more complex peripheral behaviors
by incorporating richer execution context into it (Section~\ref{sec:policies}).
The cache is hit only if the execution context matches.
In essence, the upgraded matching rule helps provide accurate responses
that reflect the specific execution context, but it sacrifices generality.

When the current execution state is detected invalid,
the symbolic execution engine switches to another branch target
and updates the matching rule and the corresponding cache entries in \kb.
If both branch targets lead to an invalid
execution state, our algorithm rolls back to the parent branch
and continues with unexplored targets (Section~\ref{sec:alg}).
We follow a depth-first-search (DFS) algorithm in the exploration.
This is because the firmware usually enters an invalid state very soon after reading an
incorrect \condr value. With DFS, we can quickly recover and switch to the right branch.
Our algorithm runs until the firmware exits (which rarely happens)
or no new basic block can be observed for a quite long time.

Although \sys follows the DFS algorithm to explore one valid path, it does not
mean dynamic analysis can only work on this path. In fact, as 
discussed in Section 1, the knowledge base built 
in the knowledge extraction phase can be used to 
emulate multiple valid paths  
in the dynamic analysis phase.  
Besides, our \kb can be dynamically enriched when the execution
meets a new peripheral register or a new execution context of existing
peripheral registers.

\paragraph{Dynamic Analysis Phase.}
Leveraging the \kb, \sys facilitates efficient dynamic analysis of
firmware by allowing arbitrary firmware to be emulated.
When a register of a custom-made peripheral is accessed,
the \kb is referred and an appropriate response value is returned and fed to
the emulation.
To demonstrate the application of this emulation capability in bug hunting, we
incorporated AFL~\cite{afl}, a popular fuzzing tool, to \sys (Section~\ref{sec:fuzzing}). 
In our prototype, we channeled the test-cases generated by AFL
to the identified data registers to fuzz the task code.
In addition,
our design is not specific to AFL and any other fuzzing tools can be used as a drop-in replacement.


\vspace*{-3mm}
\subsection{A Running Example}
\vspace*{-1.5mm}

We show a running example of the proposed approach in
Figure~\ref{fig:overview}. 
On the left, we show three execution traces on a firmware image.
A branch is represented by a node, which
is marked with the address of the peripheral register
that determines the corresponding branch targets.
In the example, two branches both correspond to
reading the peripheral register mapped at \texttt{0x40064006} at
PC \texttt{0x1a9a}.
After the knowledge extraction
phase, our algorithm decides that the third trace is valid, and the
corresponding \kb should be used in the firmware analysis phase.

In what follows, we explain how the third trace is selected and how its \kb is constructed. 
At the first branch, 
the left-side target is selected by default.
The solver calculates a value \texttt{0x30} that can lead execution
to that target.
This value is recorded as Entry 1 in the \kb for trace 1 (step 1).
The entry states that if the peripheral register
at \texttt{0x40064006} is accessed at PC \texttt{0x1a9a} later, \texttt{0x30} should be used to decide a favorable branch target.
This caching rule is encoded by the \texttt{T1} label.
Along the trace 1, the symbolic execution engine
finds that the execution state is invalid because it meets one of the rejecting conditions (see Section~\ref{sec:rejection}).
Therefore, it switches to trace 2 (step 2).
Correspondingly, Entry 1 is calculated for trace 2.
At this time, the cached value is \texttt{0x0} (step 3).
Using this value, the symbolic execution engine finds
that the left branch target is favorable at branch 2 and should be taken (step 4).
However, the execution state is proven wrong again and
the execution switches to trace 3 (step 5).
Since trace 3 is forked from trace 2,
its \kb is inherited.
However, to reach the right target at branch 2,
the symbolic execution engine finds that value \texttt{0x20}
should be used, which conflicts with Entry 1.
Therefore, the caching rule is upgraded to \texttt{T2}.
Compared with \texttt{T1}, \texttt{T2} considers
the specific execution context when a peripheral register is read, which is encoded as a hash value in the entry (step 6).
As a result, two entries of type \texttt{T2} are created, one for each branch.
In the dynamic analysis phase, which is shown on the right
part of Figure~\ref{fig:overview},
\sys queries the \kb of peripheral register access and tries to
match any entries in the \kb (and calculate the hash of execution context if necessary).
This \kb keeps \sys in valid traces.

\vspace*{-4mm}
\section{System Design \& Implementation}
\vspace*{-2mm}

We first describe the system architecture of \sys (Section~\ref{sec:armsupport}).
Then we elaborate the design and implement of \kb cache strategy (Section~\ref{sec:policies}), invalid states detection (Section~\ref{sec:rejection}), invalidity-guided \kb extraction algorithm (Section~\ref{sec:alg}), and
interrupt handling (Section~\ref{sec:interrupt}).
Finally, we describe how we integrated \sys with AFL (Section~\ref{sec:fuzzing}).

\vspace*{-3mm}
\subsection{\sys Framework}

\begin{figure}[t]
\centering
\includegraphics[width=.9\columnwidth]{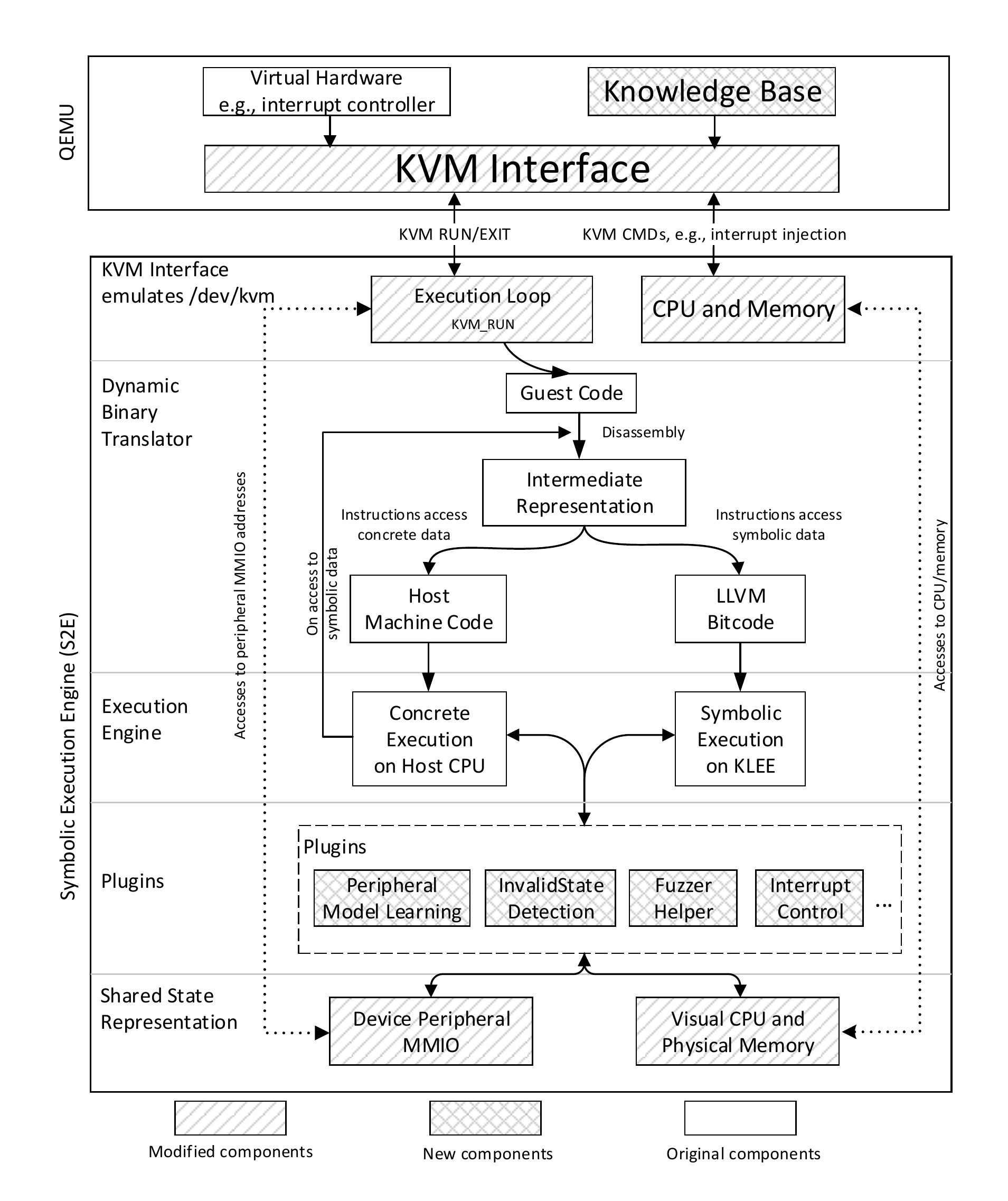}
\caption{Architecture of \sys}
\label{fig:arch}
\vspace*{-3mm}
\end{figure}

\label{sec:armsupport}
\sys is designed and developed based on S2E version 2.0, a QEMU-based concolic
execution tool for program analysis (an architecture overview of \sys in shown
in Figure~\ref{fig:arch}). As mentioned in Section~\ref{sec:s2e}, S2E provides
tens of useful plugins and APIs for analysts to use for customized analysis.
Therefore, major functions of \sys were developed as plugins to S2E using the
provided API.

Due to the aforementioned code reconstruction in S2E 2.0, the ARM support has
been dropped~\cite{armsupport}. With this release, S2E completely switched to
the KVM interface to decouple the hypervisor from the core symbolic execution
engine. Although the benefit of switching to the KVM interface is obvious, it
sacrifices broad architecture support because not every architecture can be
easily managed by the KVM interface. Particularly, ARM MCUs exhibit some
specifics making them incompatible with the canonical KVM interface. 


We made two contributions in adding ARM support to S2E. First, we ported the
DBT for ARM to S2E CPU emulation so as to emulate ARM MCUs. This task is
relatively straightforward because the upstream QEMU already supports the ARM
architecture, including ARM \cm series MCUs. We directly extracted the
corresponding logic implemented in QEMU that decodes the ARM instruction and
further interfaced it with the TCG front-end compiler. Due to the nature of
intermediate representation, the back-end of TCG was largely untouched. Then
we made necessary modifications to facilitate the communication with the core
S2E logic and to generate events that are used by the callback functions in
the S2E plugin framework. These are essential for \sys to  place hooks at
translation block boundaries and other interesting execution points.

The second task is to make the emulated ARM \cm CPU
accessible via the KVM interface.
In essence, S2E provides a virtual CPU (vCPU) capable of symbolic execution, and QEMU manages the vCPU via KVM interfaces.
Except for the canonical KVM interfaces (e.g., \texttt{KVM\_CREATE\_VCPU} to allocate a vCPU instance), 
ARM \cm CPUs exhibit many specifics that
render the implementation more challenging.
We added several customized interfaces
for QEMU to fully manage the ARM \cm vCPU via the KVM interfaces.

We developed four 
custom-made plugins to implement the designed functions in \sys:
the \texttt{InvalidStateDetection} plugin for invalid state detection (Section~\ref{sec:rejection}),
the \texttt{KnowledgeExtraction} plugin for invalidity-guided \kb extraction
and firmware emulation (Section~\ref{sec:alg}),
the \texttt{InterruptControl} plugin for interrupt injection (Section~\ref{sec:interrupt}),
and the \texttt{FuzzerHelper} plugin for fuzzer integration (Section~\ref{sec:fuzzing}).
In total,
we contributed more than 800 lines of C code to
extend S2E with ARM \cm support.
The four plugins
are completed with 829, 3,395, 311, and 560 lines of C++ code, respectively.

\vspace*{-4mm}
\subsection{\kb Caching Strategy}
\vspace*{-1mm}
\label{sec:policies}

In \sys, we use a tiered caching strategy aiming to capture
both static and dynamic behaviors of peripherals.
Specifically, four matching
rules are defined and selected adaptively based on the concrete execution context to handle the diverse complexity of
real-world firmware. 


   




\vspace*{-4mm}
\subsubsection{\twrite~--~Storage Model}
\vspace*{-2mm}

Strictly speaking, \twrite is not a matching rule.
Rather, it models the simple storage model of
peripheral registers.
That is, the peripheral register stores the
most recent value written to it and responds to the following read operations with it,
exactly as the way normal memory works.
This behavior is quite common in MCUs.
For example, the firmware writes control
values to configuration registers, which when accessed,
should respond the same value to the firmware.
\twrite is activated before any other caching rules,
provided that there was a write operation to the
register before.
When \twrite is proven wrong, it is upgraded to the caching rule \tone.





\vspace*{-3mm}
\subsubsection{\tone~--~PC-based Matching}
\vspace*{-1mm}

This matching rule reflects the greedy nature of the proposed algorithm.
It is designed to match broader peripheral accesses,
thus avoiding the path explosion issue.
To this end, it does not match specific execution context
to maximize applicability.
Specifically, the PC (\texttt{pc}) and the
peripheral address (\texttt{addr}) uniquely determine the cached value. The corresponding entry in the \kb is
encoded as \texttt{T1\_addr\_pc\_NULL\_value}.
For example, \texttt{T1\_0x40023800\_0x10000\_NULL\_0x00}
specifies that when the firmware reads from address \texttt{0x40023800} at PC \texttt{0x10000}, the value
\texttt{0x00} should be used to decide the favorable branch
target.
Based on our observation, many peripheral registers have a fixed value at a particular PC or even arbitrary PCs.
Therefore, the \tone cache rule comprises most entries for \condrs in the \kb (see Table~\ref{tab:kbentries}).
For example, in the code snippet shown in
Listing~\ref{lst:t1}, the peripheral register at \texttt{0x40023800} should always have the 17th bit set to break the while loop.
Other values are invalid and never used in the firmware.
When \tone is proven wrong, it is upgraded to the caching rule \ttwo.

\begin{lstlisting}[style=CStyle,label={lst:t1},xleftmargin=1em,framexleftmargin=1em,frame=shadowbox,caption={Code snippet of Oscillator configuration function.}]
while(MEMORY[0x40023800] & 0x20000)
	if(HAL_GetTick() >= timeout)
		return 3;
\end{lstlisting}
\vspace*{-1mm}

\vspace*{-3mm}
\subsubsection{\ttwo~--~Context-based Matching}
\vspace*{-1mm}

The \tone matching rule cannot handle complex situations 
where the returned value of the same peripheral register should change with the execution context. In Listing~\ref{lst:t2}, we show such an
example.

\begin{lstlisting}[style=CStyle,label={lst:t2},xleftmargin=1em,framexleftmargin=1em,frame=shadowbox,caption={Code snippet of UART transmission
in STM32 MCUs.}]
while(huart->TxXferCount){
	...
	if(UART_WaitOnFlagUntilTimeout(huart, 0x80, 0, tickstart, Timeout) != HAL_OK)
		return HAL_TIMEOUT;
	huart->Instance->DR = *pDataa++;
}
if(UART_WaitOnFlagUntilTimeout(huart, 0x40, 0, tickstart, Timeout) != HAL_OK)
	return HAL_TIMEOUT;
\end{lstlisting}
\vspace*{-1mm}

This code transfers a byte array via the UART interface. Before putting a byte
on the data register, it checks the status register regarding whether the hardware is ready (line 3). 
If it is ready, the status register should have a bit set as
indicated by the second parameter \texttt{0x80} of the function
\texttt{UART\_WaitOnFlagUntilTimeout}, which simply reads the status register
and compares it with the second parameter. After all the data have been sent,
the firmware reads the status register again
to check whether the transmission is completed
(line 7).
Similarly, the condition is indicated by the second parameter which
is \texttt{0x40}. The code can only return true if all the checks are passed.
In this example, accessing the same peripheral register (status register of
UART) at same PC (in \texttt{UART\_WaitOnFlagUntilTimeout()}) should yield different values, which cannot be handled by \tone.

To address this issue,
in addition to the current \texttt{pc} and peripheral register at \texttt{addr}, the \ttwo matching rule also compares
the execution context when the peripheral is accessed.
We calculate a hash value over the concatenation of execution context and encode it into the cache entry.
The resulting entry is expressed as \texttt{T2\_addr\_pc\_contextHash\_value}.
The execution context is defined as
up to three levels of caller PCs plus
current function arguments.
Therefore, in the example shown in Listing~\ref{lst:t2},
the second argument directly
distinguishes the two invocations to \texttt{UART\_WaitOnFlagUntilTimeout()} at line 3 and 7.

To show how the calling context differentiates the execution context,
we show another example in Listing~\ref{lst:t21}.
This function constantly polls the
current time (\texttt{cur\_time}) and then compares it with the time obtained
before (\texttt{timestart}) until the difference exceeds the maximum delay specified in the function parameter (\texttt{timeout}).
On a real device, the function
\texttt{ticker\_read()} reads from the peripheral a monotonically increasing counter.
To break the \texttt{while} loop, \texttt{cur\_time} must be equal to
or greater than \texttt{timestart} plus \texttt{timeout}.

\begin{lstlisting}[style=CStyle,label={lst:t21},xleftmargin=1em,framexleftmargin=1em,frame=shadowbox,caption={Code snippet of the wait() function.}]
int timestart = ticker_read();
do
    cur_time = ticker_read();
while ( cur_time - timestart < timeout );
\end{lstlisting}
\vspace*{-1mm}

Since the calling PCs at line 1 and 3 are different,
we can easily use the \ttwo caching rule to
distinguish the two invocations to \texttt{ticker\_read()}.
When \ttwo is proven wrong, it is upgraded to the caching rule \tthree.



\vspace*{-2mm}
\subsubsection{\tthree~--~Replay-based Matching}
\label{sec:t3}
\vspace*{-1mm}

However, we find that there are still corner cases which
\ttwo cannot handle.
This is particularly disconcerting when the corresponding code is related to
device initialization, since the device will not boot.
Such an example is shown in Listing~\ref{lst:t3}. 


\begin{lstlisting}[style=CStyle,label={lst:t3},xleftmargin=1em,framexleftmargin=1em,frame=shadowbox,caption={Code snippet of RF configuration.}]
rf_read_buf(&buf, len);
if (strncmp((const char*)&buf, "OK\r\n", 4))
    while (1);
\end{lstlisting}
\vspace*{-1mm}

In this MCU, the RF function is implemented on top of the UART interface.
Specifically, the data input channel of UART is used as
the control channel of the RF configuration.
When the RF module has been properly initialized,
the same UART data channel is re-purposed for RF communication.
In the code snippet,
the function \texttt{rf\_read\_buf()} reads four bytes
from the UART data register.
The result must match the string literal
``OK\textbackslash r\textbackslash n'' to conform to the
RF control protocol.
The \ttwo caching rule
cannot distinguish the four read operations to the UART data
register, since their execution contexts are exactly
the same.
When the caching rule is upgraded to \tthree,
instead of caching a single reading,
each cache entry is associated with an array of readings.
In the example,
when \sys finds the path to pass
the \texttt{strncmp} check at line 2,
four symbols obtained from \texttt{rf\_read\_buf()}
are solved together to obtain the ``OK\textbackslash r\textbackslash n''
string literal and the results are
stored in the cache entry.
Therefore, the \tthree caching rule
is encoded as 
\texttt{T3\_addr\_pc\_null\_\{v1,v2,...\}}.

In the firmware dynamic analysis phase,
the values in the array are replayed in order,
so that the execution will follow the same flow.
Therefore, it is the most specific to firmware but
is able to capture arbitrary firmware behaviors.

Based on our evaluation, the \tthree caching
rule is rarely activated.
When it is activated, most likely the corresponding
register is used for receiving external data,
as explained in the aforementioned example.
Therefore, in the firmware dynamic analysis phase,
we treat registers of type \tthree as one kind of fuzzing input points,
after replaying all the cached readings in the array.

\vspace*{-2mm}
\subsection{Invalid Execution State Detection}
\vspace*{-1mm}
\label{sec:rejection}


As mentioned before, \sys learns appropriate cache values through invalidity-guided exploration.
It is based on the assumption that during normal execution, a properly programmed
firmware should never run into any invalid states.
If an invalid state is detected, one or more of previously cached
values in the \kb should be wrong. 
In this section, we define invalid states and the
rationales behind them.
In addition, we also detail how the \texttt{InvalidStateDetection} plugin
identifies invalid states.
If an invalid state is detected by the \texttt{InvalidStateDetection} plugin, it notifies the \texttt{KnowledgeExtraction} plugin.

\paragraph{Infinite Loop.}
Typically, if the firmware
execution encounters an unrecoverable error, it will
halt itself by running a simple infinite loop.
If an infinite loop is detected, there should be a wrongly
cached peripheral reading.


The plugin keeps records of the control
flow for each execution path. If it observes repeated
cycles in the control flow, a loop is detected. To further
confirm an infinite loop, the plugin also makes sure that
the processor registers are the same in each loop.
If a register contains symbolic values, \sys solves them to concrete ones
and makes the comparison.
\sys only monitors infinite loops that occurred within
the last few translation blocks.
This number is denoted as \texttt{BB\#\_INV1} and
the default value is 30 based on our empirical study.
\texttt{BB\#\_INV1} cannot be too large for two reasons.
First, monitoring a long control flow history is time-consuming.
Second, it could mistakenly recognize the main logic of the firmware as
invalid, because the main logic of the firmware is indeed
implemented in an infinite main loop.
Fortunately, the length of the repetend in the main loop 
is often much larger than that in an invalid infinite loop.
Setting \texttt{BB\#\_INV1} to
30 effectively separates them.
In addition, infinite loop detection is only activated
when there are at least one symbol involved in the context.
The idle thread, which is typically implemented as an infinite loop
in MCU OSs, never triggers a positive infinite loop detection.

\paragraph{Long Loop.}
It is also common that the firmware waits for a certain value
in a peripheral register.
This value indicates that the peripheral has finished certain operations.
This kind of wait operation is often accompanied by a timeout mechanism, as exemplified in Listing~\ref{lst:t1}.
If \sys does not cache a correct value for this register,
there will be a long loop, taking tens of seconds to complete.

To identify a long loop, 
the \texttt{InvalidStateDetection} 
uses the same strategy to detect loops as
is done in infinite loop.
It also
counts the number of repeated cycles.
If it exceeds an adjustable value,
the plugin confirms a long loop.
The adjustable value is denoted as \texttt{BB\#\_INV2} and
we set it as 2,000 by default based on our empirical study.
Long loop detection is  only activated
when there is at least one symbol involved in the context.
Therefore, Libc functions such
as \texttt{memcpy} and \texttt{memset} never trigger
a positive long loop detection.


\paragraph{Invalid Memory Access.}
Invalid memory regions are those not mapped in the address space.
Mapped regions include ROM,
RAM, system regions and external peripheral regions.
All other are ummapped.
If the firmware accesses an unmapped memory address,
two reasons are possible.
First, the firmware itself is buggy and would encounter a memory error
even on the real device.
We consider it unlikely to happen and we did not observe
this in all the tested samples.
Second, \sys might learn a wrong response for the peripheral read operation.
The \texttt{InvalidStateDetection} plugin will report an invalid state if this happens.




\paragraph{User-defined Invalid Program Points.}
Finally, if an analyst has obtained some prior knowledge about the
firmware via static analysis or \sys itself,
we provide an interface allowing him to manually
configure additional invalid points. 
This mechanism is useful since analysts have
the option to fine-tune
the extracted knowledge about the firmware,
boosting emulation efficiency.
For example, an execution point that should never be executed (\eg~failed assertion) can be explicitly specified
by the analysts.

\SetKwRepeat{Do}{do}{while}
\begin{algorithm}[t!]
\caption{Algorithm for automatic KB extraction, denoted as KB\_Learn().}
\label{alg:learning}
\SetAlgoLined
\SetKwInOut{Input}{Input}
\SetKwInOut{Output}{Output}
\Input{KB}
\Input{selected\_target}
\Output{KB}
    symbol $\leftarrow$ get\_symbol()\;
    KB\_Update(\kb, symbol)\;
	\Do{true}{
	    targets[] $\leftarrow$ execute\_BB(selected\_target)\;
	    
		\If{meet termination condition} {
			\Return  KB\;
		}
		\If{current state is invalid} {
			break\;
		}
		\eIf{sizeof(targets) == 1}{
		    selected\_target $\leftarrow$ next\_BB(selected\_target)\;
		}{
		    selected\_target $\leftarrow$ favorable\_target(targets)\;
		    other\_target $\leftarrow$ non\_favorable\_target(targets)\;
            unexplored.push(other\_target)\;
            symbol $\leftarrow$ get\_symbol()\;
            KB\_Update(KB, symbol)\;
		}
	}
\tcp{switch execution state}
	selected\_target $\leftarrow$ unexplored.pop()\;
	KB\_Learn(selected\_target)\;
\end{algorithm}
\vspace*{-2mm}


\vspace*{-1mm}
\subsection{Invalidity-guided \kb Extraction}
\vspace*{-1mm}
\label{sec:alg}
\label{sec:termination}


In this section, we depict the proposed
knowledge base extraction algorithm for automatic peripheral modeling.

\paragraph{Branch Target Selection and Switch Algorithm.}
The algorithm, shown in Algorithm~\ref{alg:learning} and denoted as \texttt{KB\_Learn()},
is based on DFS.
It takes a basic block and the current \kb (empty for first round) as inputs, and then symbolically
executes from there.
The initial input to the algorithm is the
entry point of the firmware, which
is typically the reset handler. 
The output is the updated \kb after this round of learning.



The algorithm starts from a given branch target.
The firmware would then read a register of an unknown peripheral.
\sys assigns a symbol to it and continues execution until a branch is met.
The algorithm gets the symbol responsible for the
branch target and then updates the \kb using 
the algorithm listed in Algorithm~\ref{alg:updateKB}, which we explain later.
The main body is a while loop to step over basic blocks.
After finishing each basic block, it checks
if the current execution state meets the
termination conditions (explained later).
If so, the algorithm returns the current \kb.
If no termination condition is met, it then checks if the current execution state is valid or not
based on the conditions mentioned in Section~\ref{sec:rejection}.
If the state is valid, 
it judges if a branch is reached.
If a branch is not reached, the next basic block is selected to continue the while loop.
If a branch is reached, the algorithm selects a favorable target
according to the existing \kb and sets it as the next branch target.
The non-favorable target is pushed back to a stack for future exploration.
Then,
the algorithm gets the symbols responsible for the
favorable branch target and updates the \kb.
The only condition to break the while loop is that
an invalid execution state is detected in line 8.
If this happens,
the next branch target is popped from the stack, and the
algorithm recursively executes from there.

\begin{algorithm}[t]
\caption{Algorithm for updating the knowledge base, denoted as KB\_Update()}
\label{alg:updateKB}
\SetAlgoLined
\SetKwInOut{Input}{Input}
\Input{KB}
\Input{symbol}
new\_entry $\leftarrow$ solver(symbol)\;
\eIf{new\_entry conflicts with KB} {
	\tcp{upgrade caching rules}
\uIf{type(symbol) == T0}{
    type(symbol) $\leftarrow$ T1\;}
\uElseIf{type(symbol) == T1}{
type(symbol) $\leftarrow$ T2\;
}
\ElseIf{type(symbol) == T2}{
type(symbol) $\leftarrow$ T3\;
}
    replace the conflicting entry with new\_entry\;
}{
KB $\leftarrow$ KB | new\_entry\;
}
\end{algorithm}
\vspace*{-1mm}

\paragraph{\kb Update Algorithm.}
Next, we explain the knowledge base update algorithm shown in
Algorithm~\ref{alg:updateKB}, denoted as \texttt{KB\_Update()}.
It takes the current \kb and a symbol as inputs.
First, the symbolic execution engine solves a concrete value
for the symbol that could lead the execution to the current branch target.
The returned concrete value is used to construct a new cache entry.
If the new entry does not conflict with the current \kb,
it is inserted to the \kb.
Otherwise, the caching rule of the corresponding symbol is upgraded.
Specifically, \twrite is upgraded to \tone; \tone is upgraded
to \ttwo and so forth.

\paragraph{Termination Condition.}
Real-world firmware typically runs 
in an infinite loop to respond to external events,
therefore would never exit.
Therefore a round of knowledge extraction could last forever.
In our prototype, we monitor the lastly executed 30,000 basic blocks
and make sure that no new basic blocks are reached.
If this happens, this round of knowledge extraction terminates.
Note the number of monitored basic blocks is an empirical value and can be adjusted by changing \texttt{BB\#\_Term}.


\paragraph{Reinforced Learning.}
To emulate a firmware image, \sys starts execution from the entry point
following \texttt{KB\_Learn()}. The first round usually takes a long time
since the \kb has not been set up. As more cache entries are being built,
accesses to peripheral registers lead to more cache hits. However, one-shot
knowledge extraction cannot guarantee full coverage of all peripherals,
especially considering that many hard-to-reach code regions are only executed
when specific events happen. If we find that the current \kb does not include
cache entries for certain registers or the context hash/PC cannot be matched,
\sys needs to conduct another round of knowledge extraction phase to learn
additional peripheral behaviors. We call it reinforced learning. In a real
firmware emulation, multiple rounds are needed when new peripherals are
discovered by new test-cases.

\vspace*{-4mm}
\subsection{Interrupt Handling}
\vspace*{-1mm}
\label{sec:interrupt}
The interrupt is important for peripheral to interact with
the external world. Without interrupts, many firmware
behaviors cannot be triggered.

\vspace*{-2.5mm}
\subsubsection{Interrupt Delivery}
\vspace*{-1mm}
Although QEMU has implemented
a virtual interrupt controller (\ie~NVIC) for ARM \cm MCUs,
which could be used to dispatch and respond to interrupts,
this function is largely limited to implementing
system peripherals such as \textit{SYSTICK},
because QEMU does not know when to fire interrupts for
custom-made peripherals.
First,
to find out which interrupt is activated,
the \texttt{InterruptControl} plugin checks the NVIC
\textit{Interrupt Set Enable Register} (ISER).
Then, for deterministic replay of interrupt sequences,
our prototype follows a similar interrupt firing strategy as \ppim.
The plugin
delivers activated interrupts (via setting 
the corresponding
bit of the NVIC \textit{Interrupt Status Pending Register} (ISPR))
in a round-robin fashion at a
fixed interval defined by the user.
As empirical values, in our evaluation,
we set the interval to be once every 2,000 basic blocks
during the knowledge extraction phrase and once every 1,000 basic blocks during the analysis phase.

\vspace*{-2mm}
\subsubsection{Caching Strategy for Interrupts}
\vspace*{-1.5mm}

Using the caching strategy explained in Section~\ref{sec:policies},
we found that the code coverage inside
the interrupt handler is severely limited.
It turned out our algorithm over-approximates the paths.
Normally, the interrupt handler of a peripheral often
executes different paths based on the values of the
control register and status register.
All these paths are valid from the viewpoint of our invalidity checking mechanism.
Unfortunately, with the cache mechanism, only one path can be executed.
An example is shown in Listing~\ref{lst:tirq},
in which the UART driver decides to invoke the receive or transmit
function based on the value of the status register \texttt{isrflags} and
control register \texttt{crlflags}.
When these registers have cache entries
in the \kb, the emulated path would be fixed.



\begin{lstlisting}[style=CStyle,label={lst:tirq},xleftmargin=1em,framexleftmargin=1em,frame=shadowbox,caption={Code snippet of the UART interrupt handler in the STM32 HAL library.}]
void UART_IRQHandler(UART_Handle *huart) {
    uint32_t isrflags = READ_REG(huart->SR);
    uint32_t cr1flags = READ_REG(huart->CR1);
    /* UART in Receiver mode */
    if(((isrflag & USART_SR_RXNE) != RESET)
    && ((crlflag & USART_CR1_RXNEIE) != RESET)){
        UART_Receive_IT(huart);
        return;
    }
    /* UART in Transmitter mode */
    if(((isrflag & USART_SR_TXE) != RESET)
     && ((crlflag & USART_CR1_TXEIE) != RESET)) {
        UART_Transmit_IT(huart);
        return;
    }
    ...
}
\end{lstlisting}
\vspace*{-1mm}


To solve this problem, \sys tries to execute different paths in an interrupt handler.
Specifically, \sys monitors the execution context.
If the interrupt context is detected,
the symbolic execution engine tries to explore all the possible paths.
The readings for a peripheral register that lead
to all valid execution states are collectively stored
in the corresponding cache entry.
In the firmware dynamic analysis phase, the values in each
entry are randomly selected. As such, paths in an interrupt
handler will be randomly executed.

However,
it usually takes multiple tries before triggering the intended interrupt event.
We rely on an observation to
increase the accuracy of interrupt event prediction.
Specifically,
in peripherals, the status registers are often dependent on the control registers and
thus can be ignored in condition statements.
Moreover, control registers are typically recognized as \twrite,
so we can accurately infer their values by 
referring to the most recent written values.
Therefore, \sys looks for peripheral registers
of type \twrite at first.
If it is found,
\sys uses the most recently written value to it to
calculate the branch target, regardless of whether other registers are also
involved in the condition statement. 
If it is not found, \sys randomly selects
all possible values of type \tone and \ttwo to drive the execution.
This optimization helps \sys accurately handle many common peripherals such as UART and I2C.
\vspace*{-3mm}
\subsection{Fuzzer Integration}
\vspace*{-2mm}
\label{sec:fuzzing}

The
\texttt{FuzzerHelper} plugin is used to accommodate AFL so that it can be bridged to \sys. Also, it automatically
finds fuzzing input points to feed data to the tested tasks.

\paragraph{AFL Accommodation.}
Although
AFL already supports fuzzing binaries running QEMU,
it is limited to fuzzing user-space binaries.
As such,
we only use AFL for test-case generation and
leave the rest to \texttt{FuzzerHelper}, including the coverage instrumentation, fork sever, and crash/hang detection.
This also allows us to readily replace AFL for alternative fuzzing tools
with minimal re-engineering effort.
We implemented the same path coverage algorithm with the AFL.
Concretely, the code coverage information is collected
by tracking the translation block transitions.
Then, we share the bitmap of code coverage information with AFL via shared memory.
For the fork sever, we consider the moment the firmware reads the 
first byte of test-case as the fork point. 
We used the existing interface \texttt{forkAndConcretize} in S2E
to take a snapshot of the whole execution state when the execution
reaches the fork point for the first time.
We choose the default fork point as the program point at which the firmware reads a data register for the first time.
Then, every time the execution finishes reading test-case or
the firmware crashes/hangs, 
the plugin rolls back to the fork point and clones
another state to continue fuzzing.
For crash detection, 
we implemented a very basic memory error detector, which checks the
memory access permissions based on regions: R+X for the whole ROM,
R+W for RAM, peripherals, and system control block, and
no access for the rest.
We also consider \texttt{HardFault} as a crash indicator
because typically it means an unrecoverable error.
The timeout is set as 10 seconds for hang detection.

\paragraph{Data Registers Identification.}
In fuzz testing, it is essential to identify input channels under attackers' control.
In MCUs, this corresponds to peripheral data registers.
We found candidate data registers often exhibit the following characteristics, which gave us opportunities to
identify them automatically. 
First, the \tthree registers are mostly data registers.
This is because the readings from them are often protocol data,
as exemplified in Listing~\ref{lst:t3}.
Second, data registers are often read in interrupt handlers but their readings are consumed in the non-interrupt context.
Third, compared
with other kinds of registers,
data registers are frequently accessed during execution (more than hundreds of times).
If a register has one of the above characteristics,
we mark it as a data register for fuzzing.
As shown in Table~\ref{tab:configuration}, this method enables us to accurately identify data registers
for real-world fuzzing.

\vspace*{-2mm}
\section{Evaluation}
\vspace*{-2mm}
The main evaluation questions for \sys are as follows.
1) whether it is able to emulate the behaviors of different kinds of unknown peripherals
correctly;
2) whether the performance is within an acceptable range for practical uses;
3) whether it enables analysis tools like fuzzers
to find real-world bugs of the task code of firmware.
All experiments were conducted
on an 8-core/16-thread Xeon server with 48GB RAM, running a Ubuntu 18.04 OS.


\vspace*{-3mm}
\subsection{Unit Tests}
\vspace*{-1mm}
\label{sec:unittest}

We conducted the same unit-test experiment as
was done in \ppim to ensure a head-to-head comparison.
It tests how \sys can handle individual peripheral functions.


\vspace*{-3mm}
\subsubsection{Experimental Setup}
\vspace*{-1mm}
We reused the same 66 firmware samples in the \ppim experiment~\cite{p2imgit}.
These samples cover eight most popular MCU peripherals, three MCU chips (STM32 F103RB, NXP MK64FN1M0VLL12, and Atmel SAM3X8E),
and three widely used MCU OS/system libraries (NuttX, RIOT, and Arduino).
Each unit-test sample represents a unique and feasible combination of a
peripheral, an OS, and an SoC. After rebooting,
the firmware performs the basic peripheral operations.
For each unit test, we first ran the knowledge extraction phase.
During dynamic analysis,
we overrode the testcases generated by AFL with the expected data extracted
from the unit test to emulate data input.


\vspace*{-3mm}
\subsubsection{Experiment Results}
\vspace*{-1mm}
The results are summarized in Table~\ref{tab:unittest}.
All the unit tests finished the knowledge extraction phase within one minutes with one round.
It suggests the high efficiency of our knowledge extraction algorithm.
Out of 66 samples, only three unit tests failed, suggesting a passing rate of 95\%,
which is higher than the result in \ppim (79\%).

\begin{table*}[t]
\caption{Unit tests results without human intervention}
\centering
\begin{adjustbox}{width=\textwidth}
\begin{tabular}{l|l|lll|l|ll}
\textbf{Peripheral}  & \textbf{Functional Operations}       & \textbf{F103/Arduino} & \textbf{F103/RIOT} & \textbf{F103/NUTTX} & \textbf{K64F/RIOT} & \textbf{SAM3/Arduino} & \textbf{SAM3/RIOT} \\ \hline
\textbf{ADC}                    & Read an analog-to-digital conversion           & Pass                  & N/A                & Pass                & Pass               & Pass                  & Pass               \\\hline
\textbf{DAC}                    & Write a value for digital-to-analog conversion & N/A                   & N/A                & N/A                 & N/A                & Pass                  & Pass               \\\hline
\multirow{3}{*}{\textbf{GPIO}}  & Execute callback after pin interrupt           & Pass                  & Pass               & Pass                & Pass               & Pass                  & Pass               \\
                                & Read status of a pin                           & Pass                  & Pass               & Pass                & Pass               & Pass                  & Pass               \\
                                & Set/Clear a pin                                & Pass                  & Pass               & Pass                & Pass               & Pass                  & Pass               \\\hline
\textbf{PWM}                    & Configure PWM as an autonomous peripheral      & Pass                  & N/A                & Pass                & Pass                & Pass                  & Pass               \\\hline
\multirow{2}{*}{\textbf{I2C}}   & Read a byte from a slave                       & Pass                  & N/A*               & \textbf{Fail}               & \textbf{Fail}               & Pass                  & N/A                \\
                                & Write a byte to a slave                        & Pass                  & N/A*               & N/A*               & \textbf{Fail}               & Pass                  & N/A                \\\hline
\multirow{2}{*}{\textbf{UART}}  & Receive a byte                                 & Pass                  & Pass               & Pass                & Pass               & Pass                  & Pass               \\
                                & Transmit a byte                                & Pass                  & Pass               & Pass                & Pass               & Pass                  & Pass               \\\hline
\multirow{2}{*}{\textbf{SPI}}   & Receive a byte                                 & Pass                  & Pass               & Pass                & Pass               & Pass                  & Pass               \\
                                & Transmit a byte                                & Pass                  & Pass               & N/A*                & Pass               & Pass                  & Pass               \\\hline
\multirow{2}{*}{\textbf{Timer}} & Execute callback after interrupt               & N/A                   & Pass               & N/A                 & Pass               & N/A                   & Pass               \\
                                & Read counter value                             & N/A                   & Pass               & N/A                 & Pass               & N/A                   & Pass              
\end{tabular}

\end{adjustbox}
\label{tab:unittest}
\flushleft
\scriptsize{
Note: 1. There are 18 unavailable entries (marked with ``N/A'') 
because these combinations of MCU Soc and OS/libraries are not correctly supported by real devices. The original \ppim paper marked 14 of them.
There are 4 additional ones (marked with ``*'') after we confirmed with the \ppim authors.

\quad \quad ~ 2. Since unit test-cases are simple, we set BB\#\_INV1 as 15, BB\#\_INV2 as 500 and BB\#\_Termination as 10,000 for all unit test samples. 
}
\vspace*{-3mm}
\end{table*}


\paragraph{Failed Tests in \ppim.}
A major reason for failed tests in \ppim is register mis-categorization.
When the register is treated as another type, the resulting response is very likely to be wrong.
We attribute failed tests to several reasons, including
mis-categorization (MC), invalid assumption (IA) and limited exploration (LE),
which are explained in Section~\ref{sec:ppimfailreasons}.

\paragraph{Failed Tests in \sys.}
Invalidity checking plays an important role in \sys. 
If an unexpected path is not recognized as invalid, \sys may lead the emulation to it.
The failed tests were all caused by this issue.
In Listing~\ref{lst:i2ctestsample}, we show such an example in which
the firmware reads a byte via the I2C bus.
It first checks the status register.
If an error condition is detected in line 3, the function returns an error.
Otherwise, the normal function is performed.

\begin{lstlisting}[style=CStyle,label={lst:i2ctestsample},xleftmargin=1em,framexleftmargin=1em,frame=shadowbox,caption={Code snippet in which \sys fails to extract correct information.}]
int i2c_read_bytes(...){
    I2C_TypeDef *i2c_dev = i2c_config[dev].dev;
    if ((i2c_dev->SR & 6) == 2)
        return Error;
    ...
    data = i2c_dev->DR;
}
\end{lstlisting}
\vspace*{-2mm}

In this example, the error returned in line 5 is not handled.
As a result, regardless of the path being executed in the function,
the execution error cannot be detected by the proposed invalidity checking mechanism.
In our evaluation, three out of 66 test-cases have this issue.
We argue that this problem is mainly due to not following
the best practice in programming.
In particular, well implemented firmware should detect the error code and handle it immediately.
This problem can also be mitigated by invoking the provided
interface to specify invalid program points.
In this example, line 4 should be avoided.
Therefore, the analyst can configure the address of line 4 as an invalid program point, so that the
\texttt{InvalidStateDetection} plugin is able to detect it (Section~\ref{sec:rejection}) when line 4 is executed.
After adding one additional invalid point 
to each failed sample, \sys achieved a 100\% passing rate.

\vspace*{-3mm}
\subsection{Fuzzing with \sys}

\begin{table*}[t]
\caption{Results of knowledge extraction and fuzzing with \sys}
\centering
\begin{adjustbox}{width=\textwidth}
\begin{tabular}{cl|cr|rrrr|rrrl|rr}
\multicolumn{1}{l}{\textbf{}}                                                                                                        & \textbf{}                  & \multicolumn{6}{c|}{\textbf{Knowledge Extraction   Performance}}                                                                                                                                                                                                                                                                                                                      & \multicolumn{1}{l}{\textbf{}}     & \multicolumn{1}{l}{\textbf{}}       & \multicolumn{1}{l}{\textbf{}}                                                            & \textbf{}                             & \multicolumn{1}{l}{\textbf{}}                                                             & \multicolumn{1}{l}{\textbf{}}                                                           \\
\multicolumn{1}{l}{\textbf{}}                                                                                                        & \textbf{}                  & \textbf{}         & \multicolumn{1}{c|}{\textbf{}}                                                             & \multicolumn{2}{c}{\textbf{\begin{tabular}[c]{@{}c@{}}w/Cache\\      Round \#1\end{tabular}}}                                    & \multicolumn{2}{c|}{\textbf{\begin{tabular}[c]{@{}c@{}}w/o Cache\\      Round \#1\end{tabular}}}                                  & \multicolumn{4}{c|}{\textbf{Coverage Improvement*}}                                                                                                                                                         & \multicolumn{2}{c}{\textbf{Fuzzing}}                                                                                                                                                \\
\multicolumn{1}{c|}{\textbf{Refs}}                                                                                                   & \textbf{Firmware}          & \textbf{Round \#} & \multicolumn{1}{c|}{\textbf{\begin{tabular}[c]{@{}c@{}}Total\\      Time(s)\end{tabular}}} & \multicolumn{1}{c}{\textbf{\begin{tabular}[c]{@{}c@{}}Path\\      Coverage\end{tabular}}} & \multicolumn{1}{c}{\textbf{Time(s)}} & \multicolumn{1}{c}{\textbf{\begin{tabular}[c]{@{}c@{}}Path\\      Coverage\end{tabular}}} & \multicolumn{1}{c|}{\textbf{Time(s)}} & \multicolumn{1}{c}{\textbf{QEMU}} & \multicolumn{1}{c}{\textbf{w/\sys}} & \multicolumn{1}{c}{\textbf{\begin{tabular}[c]{@{}c@{}}Improv.\\      Rate\end{tabular}}} & \multicolumn{1}{c|}{\textbf{w/\ppim}} & \multicolumn{1}{c}{\textbf{\begin{tabular}[c]{@{}c@{}}Crashes\\ True/False\end{tabular}}} & \multicolumn{1}{c}{\textbf{\begin{tabular}[c]{@{}c@{}}Hangs\\ True/False\end{tabular}}} \\ \hline
\multicolumn{1}{c|}{\multirow{10}{*}{\textbf{P2IM~\cite{fengp2020p2im}}}}                                                            & \textbf{CNC}               & 2                 & 49s                                                                                        & 4/689                                                                                     & 18s                                  & 605/3080                                                                                  & 2h*                                   & 2.68\%                            & 67.96\%                             & 24.96x                                                                                   & Y(66.50\%)                            & 0/0                                                                                       & 0/0                                                                                     \\
\multicolumn{1}{c|}{}                                                                                                                & \textbf{Console}           & 1                 & 5s                                                                                         & 2/147                                                                                     & 5s                                   & 28/250                                                                                    & 31s                                   & 2.19\%                            & 35.90\%                             & 16.42x                                                                                   & Y(46.30\%)                            & 0/0                                                                                       & 0/0                                                                                     \\
\multicolumn{1}{c|}{}                                                                                                                & \textbf{Drone}             & 1                 & 593s                                                                                       & 2/412                                                                                     & 593s                                 & 167/2080                                                                                  & 2h*                                   & 8.40\%                            & 89.74\%                             & 10.69x                                                                                   & Y(74.75\%)                            & 0/0                                                                                       & 0/0                                                                                     \\
\multicolumn{1}{c|}{}                                                                                                                & \textbf{Gateway}           & 9                 & 173s                                                                                       & 5/543                                                                                     & 16s                                  & 3/364                                                                                     & 12s                                   & 1.70\%                            & 52.71\%                             & 30.94x                                                                                   & Y(54.51\%)                            & 6/0                                                                                       & 0/0                                                                                     \\
\multicolumn{1}{c|}{}                                                                                                                & \textbf{Heat\_Press}       & 1                 & 26s                                                                                        & 2/424                                                                                     & 26s                                  & 2/652                                                                                     & 673s                                  & 1.11\%                            & 30.21\%                             & 27.22x                                                                                   & Y(32.68\%)                            & 2/0                                                                                       & 0/0                                                                                     \\
\multicolumn{1}{c|}{}                                                                                                                & \textbf{PLC}               & 3                 & 33s                                                                                        & 6/143                                                                                     & 9s                                   & 4/170                                                                                     & 12s                                   & 3.51\%                            & 26.44\%                             & 7.53x                                                                                    & Y(27.43\%)                            & 139/0                                                                                     & 0/0                                                                                     \\
\multicolumn{1}{c|}{}                                                                                                                & \textbf{Reflow\_Oven}      & 2                 & 267s                                                                                       & 6/372                                                                                     & 165s                                 & 4/348                                                                                     & 36s                                   & 3.57\%                            & 40.53\%                             & 11.3x                                                                                    & Y(34.96\%)                            & 0/0                                                                                       & 0/0                                                                                     \\
\multicolumn{1}{c|}{}                                                                                                                & \textbf{Robot}             & 1                 & 53s                                                                                        & 9/437                                                                                     & 53s                                  & 676/2986                                                                                  & 2h*                                   & 2.47\%                            & 43.25\%                             & 17.51x                                                                                   & Y(46.87\%)                            & 0/0                                                                                       & 0/0                                                                                     \\
\multicolumn{1}{c|}{}                                                                                                                & \textbf{Soldering\_Iron}   & 3                 & 115s                                                                                       & 11/875                                                                                    & 44s                                  & 5/348                                                                                     & 34s                                   & 4.21\%                            & 62.01\%                             & 14.73x                                                                                   & Y(48.55\%)                            & 0/32                                                                                      & 0/4                                                                                     \\
\multicolumn{1}{c|}{}                                                                                                                & \textbf{Steering\_Control} & 1                 & 15s                                                                                        & 2/389                                                                                     & 15s                                  & 3/1275                                                                                    & 481s                                  & 0.68\%                            & 32.59\%                             & 48.09x                                                                                   & Y(29.02\%)                            & 12/0                                                                                      & 0/0                                                                                     \\ \hline
\multicolumn{1}{c|}{\multirow{2}{*}{\textbf{\begin{tabular}[c]{@{}c@{}}HALucinator\\ ~\cite{clements2020halucinator}\end{tabular}}}} & \textbf{6LoWPAN\_Sender}   & 6                 & 287s                                                                                       & 4/876                                                                                     & 88s                                  & 350/4231                                                                                  & 49m                                   & 0.88\%                            & 48.30\%                             & 55.17x                                                                                   & N(LE)                                 & 0/0                                                                                       & 0/0                                                                                     \\
\multicolumn{1}{c|}{}                                                                                                                & \textbf{6LoWPAN\_Receiver} & 6                 & 293s                                                                                       & 4/875                                                                                     & 89s                                  & 350/4232                                                                                  & 50m                                   & 0.88\%                            & 47.36\%                             & 54.08x                                                                                   & N(LE)                                 & 2/0                                                                                       & 0/0                                                                                     \\ \hline
\multicolumn{1}{c|}{\multirow{2}{*}{\textbf{Pretender~\cite{gustafson2019toward}}}}                                                  & \textbf{RF\_Door\_Lock}    & 1                 & 117s                                                                                       & 4/332                                                                                     & 117s                                 & 8/876                                                                                     & 43m                                   & 0.25\%                            & 24.37\%                             & 97.57x                                                                                   & N(MC)                                 & 98/0                                                                                      & 0/0                                                                                     \\
\multicolumn{1}{c|}{}                                                                                                                & \textbf{Thermostat}        & 2                 & 449s                                                                                       & 5/686                                                                                     & 412s                                 & 13/393                                                                                    & 2h*                                   & 0.18\%                            & 25.48\%                             & 143.85x                                                                                  & N(MC)                                 & 76/0                                                                                      & 0/0                                                                                     \\ \hline
\multicolumn{1}{c|}{\textbf{WYC~\cite{muench2018you}}}                                                                               & \textbf{XML\_Parser}       & 2                 & 54s                                                                                        & 5/572                                                                                     & 39s                                  & 339/2517                                                                                  & 106m                                  & 0.72\%                            & 26.31\%                             & 36.45x                                                                                   & N(IA)                                 & 9/0                                                                                       & 0/0                                                                                     \\ \hline
\multicolumn{1}{c|}{\multirow{6}{*}{\textbf{\sys}}}                                                                                  & \textbf{GPS\_Tracker}      & 3                 & 57s                                                                                        & 4/304                                                                                     & 22s                                  & 7/155                                                                                     & 17s                                   & 0.49\%                            & 23.90\%                             & 48.81x                                                                                   & N(MC)                                 & 0/0                                                                                       & 0/29                                                                                    \\
\multicolumn{1}{c|}{}                                                                                                                & \textbf{LiteOS\_IoT}       & 3                 & 62s                                                                                        & 13/537                                                                                    & 28s                                  & 99/2884                                                                                   & 99m                                   & 3.60\%                            & 62.33\%                             & 17.33x                                                                                   & N(MC)                                 & 0/0                                                                                       & 0/0                                                                                     \\
\multicolumn{1}{c|}{}                                                                                                                & \textbf{Zepyhr\_SocketCan} & 4                 & 535s                                                                                       & 5/1634                                                                                    & 336s                                 & 26/410                                                                                    & 45s                                   & 1.14\%                            & 47.41\%                             & 41.47x                                                                                   & N(MC,LE)                              & 0/0                                                                                       & 0/0                                                                                     \\
\multicolumn{1}{c|}{}                                                                                                                & \textbf{3Dprinter}         & 2                 & 25s                                                                                        & 4/512                                                                                     & 18s                                  & 589/1398                                                                                  & 2h*                                   & 0.50\%                            & 18.94\%                             & 38.17x                                                                                   & N(IA)                                 & 0/0                                                                                       & 0/123                                                                                   \\
\multicolumn{1}{c|}{}                                                                                                                & \textbf{\bm{$\mu$}tasker\_MODBUS}   & 4                 & 256s                                                                                       & 3/877                                                                                     & 95s                                  & 13/1236                                                                                   & 18m                                   & 0.64\%                            & 60.30\%                             & 94.25x                                                                                   & N(MC,LE)                              & 0/0                                                                                       & 0/0                                                                                     \\
\multicolumn{1}{c|}{}                                                                                                                & \textbf{\bm{$\mu$}tasker\_USB}      & 6                 & 227s                                                                                       & 2/491                                                                                     & 45s                                  & 4/342                                                                                     & 31s                                   & 0.68\%                            & 41.97\%                             & 61.95x                                                                                   & N(MC,LE)                              & 47/0                                                                                      & 0/0                                                                                    
\end{tabular}
\end{adjustbox}
\label{tab:kbandfuzz}
\flushleft
\scriptsize{
*: Coverage = \# of visited QEMU translation blocks / total \# of basic blocks. This is the same method used in \ppim. The absolute numbers can be found in Table~\ref{tab:covnum}.
}
\vspace*{-4mm}
\end{table*}

\label{sec:experiment2}

\subsubsection{Experimental Setup}
\vspace*{-1.5mm}
To comprehensively evaluate our work,
we obtained the ten firmware samples used in \ppim~\cite{p2imgit},
two used in HALucinator~\cite{halgit}, two used in  Pretender~\cite{pretednergit}, and one used in the paper WYCINWYC~\cite{muench2018you}.
In addition, we collected six extra firmware samples 
running on real-world commercial devices.
The source and a brief description for each extra firmware sample can be found
in Appendix~\ref{sec:appendix3}.
In total, our sample set includes 21 real-world firmware images.
In general, these samples collectively cover more than ten MCU models from top MCU vendors such as Atmel, NXP, Maxim, and STM by revenue~\cite{topmcu}.
Each of them includes a diverse set of peripherals, including UART, CAN, Radio, USB, etc. 
and popular OSs/libraries such as FreeRTOS, RIOT, and Arduino.
All on-chip peripherals used by each firmware is listed in Table~\ref{tab:detail}.



In the experiment, 15 samples were tested under the default
configuration without any manual inputs during KB extraction.
For the remaining six samples,
only one user-defined invalid program point (see Column 5 in Table~\ref{tab:configuration}) needs
to be added for each to enhance the invalidity checking.
Except for three samples which need analysts to specify one additional data register (\ie~bracketed registers in the last column in Table~\ref{tab:configuration}),
others directly used the automatically identified data registers.
The detailed information about the configuration for each tested sample
can be found in Table~\ref{tab:configuration} of Appendix~\ref{sec:appendix2}.


As a comparison, we used \ppim to conduct experiments on the same set of firmware samples.
To ensure a fair comparison, we strictly followed the instructions on
\ppim GitHub repo~\cite{p2imrepo} and communicated with the authors
when something uncertain was encountered.
We performed the following manual works when using \ppim.
First, for each sample, we modified the source code 
to explicitly invoke the function \texttt{startForkserver} for AFL
fuzzing integration\footnote{\url{https://github.com/RiS3-Lab/p2im/blob/master/docs/prep_fw_for_fuzzing.md}\label{fn:prefirmware}}.
Second, we manually
added new board and MCU memory regions to the \ppim source code\footnote{\url{https://github.com/RiS3-Lab/p2im/blob/master/docs/add_mcu.md}}.
Note that the same information is also needed for \sys.
However, we provided an easy-to-use Lua-based interfaces to quickly
configure the MCU without modifying the QEMU C source code.

\vspace*{-3mm}
\subsubsection{Experiment Results}
\vspace*{-2mm}
For each sample,
we first ran a round of knowledge extraction,
and then started fuzzing for 24 hours.
If reinforced knowledge extraction is triggered,
\sys automatically switches back and forth between
the knowledge extraction phase and dynamic analysis (fuzzing) phase.
We evaluated the results in three aspects.
First, we measured the total time and the number of rounds needed in \kb extraction.
We show the performance improvement with the cache mechanism.
Second, we measured the path coverage with and without \sys and
compared the result with \ppim.
Finally, we show the fuzzing results.

\paragraph{Knowledge Extraction Performance.}
We recorded the total number of rounds of reinforced learning and
the total time spent on knowledge extraction across multiple rounds.
Table~\ref{tab:kbandfuzz} shows the results.
In the worst case, the knowledge extraction phase took less than ten
minutes, while
for most samples the knowledge extraction phase can complete within two minutes.
Some complex firmware like \texttt{Gateway}
discovered multiple new peripheral registers during fuzzing
and therefore switched between the knowledge extraction phase and
the fuzzing phase back and forth several times.

The performance of knowledge extraction is good enough for practical use cases, 
especially considering that the \kb can be reused multiple times in firmware analysis.
The reason for knowledge extraction process being so efficient is attributed to the cache mechanism used in the exploration algorithm.
In the right part of the \textbf{knowledge extraction performance} column in Table~\ref{tab:kbandfuzz}, we show the number of paths being searched and consumed time in the symbolic execution
with and without using cache \kb during the knowledge extraction phase. 
For the experiments without using the cache, a target branch was randomly selected in the exploration.
As shown in the table, using the cache to select favorable branches, much less time is spent
and fewer paths need to be explored to finish a round of knowledge extraction.
Without using cache, some firmware cannot finish the first round.
In these cases, we forcedly stopped the execution after two hours.

\paragraph{Coverage Improvement.}
As shown in the Table~\ref{tab:kbandfuzz},
the code coverage increases 10x to 140x compared to
that in the normal QEMU without peripheral emulation.


\label{sec:ppimmisuse}
In the column showing the results of \ppim,
we marked a letter ``Y'' for samples that \ppim can emulate and noted the coverage in the bracket.
For those that \ppim cannot emulate, we marked a letter ``N'' and noted the reasons.
The detailed explanation for the failure reasons can be found in Section~\ref{sec:ppimfailreasons}.
We observe slight improvement in code coverage over \ppim.

\paragraph{Fuzzing.}
We used our tool to fuzz the task code in the collected samples.
These tasks take inputs from the identified data registers.
We were able to reproduce all the bugs mentioned
in previous works, except for XML parser sample in WYCNINWYC~\cite{muench2018you}.
This missed bug is caused by a heap overflow, which can only be detected
with a fine-grained memory checker such as AddressSanitizer~\cite{asan}.
Designing an advanced memory checker is orthogonal to this work.


In addition to known bugs,
we also found two previously unknown bugs in \texttt{Steering\_Control}
and \texttt{$\mu$TaskerUSB}.
The bug in \texttt{$\mu$TaskerUSB} is caused by an out-of-bound write.
The USB driver only uses a receive buffer of 512 bytes to
read an input of up to 1,024 bytes,
resulting in DoS or data corruption.
This result is encouraging because the same samples
have been extensively fuzzed in previous works,
and we can reasonably anticipate that \sys is likely to find more bugs.
The bug in \texttt{Steering\_Control}
is caused by a double-free of a string buffer, allowing for arbitrary write.
More specifically,
the firmware uses dynamic memory to store the received data from the serial port.
If the memory allocation fails, the same buffer will be freed twice.
We have reported the bugs to the corresponding device vendors.
Since \texttt{Steering\_Control} was also tested by \ppim but \ppim failed to find the bug,
we further studied the root cause.
It turned out this is due to the way it handles test-cases.
Specifically, \ppim requires the user to manually set the fork point for fuzzing. 
In this firmware configuration,
\ppim only handles very few bytes at the beginning of each test-case, whereas this bug is only triggered when a long input has been processed. We note this issue is not caused by \ppim's limited emulation capability and is fixable by specifying a proper forking point.

Since we adopted a different strategy in selecting the fork point with \ppim
(automatic vs. manual), it is unfair to directly compare the number of
executed test-cases per second to evaluate the fuzzing speed.
Instead, we measured the execution time 
to complete one million basic blocks for \sys and \ppim to evaluate the speed.
In our experiments,
we observed a slight slow-down of \sys compared with \ppim (1.2x to 1.7x).
We attribute this to the slower execution speed of S2E.
S2E introduces nearly 1.5X runtime overhead over the vanilla QEMU due to the check of symbolic data
in each translation block execution.
This problem can be alleviated using the single-path mode of S2E~\cite{s2esinglepath}.





\paragraph{False Crashes/Hangs.}
In our evaluation, we observed some false positives in \texttt{Steering\_Control}, \texttt{GPS\_Tracker} and \texttt{3Dprinter}.
After careful examination, they were caused by the lack of Direct Memory Access (DMA) support in \sys.
DMA allows the peripherals to directly access the RAM
independent of the processor.
Since it is not simply responding values to peripheral access operations,
symbolic execution cannot provide any useful knowledge.
A recent work~\cite{mera2020dice} has been specifically designed to handle DMA.

\vspace*{-3mm}
\subsection{Failure Reasons in \ppim}
\vspace*{-1mm}
\label{sec:ppimfailreasons}

This section explains the root causes for failed emulations in \ppim.
We use the same notation as Table~\ref{tab:kbandfuzz} to refer to
the causes.

\paragraph{MC~--~Mis-categorization of Registers.}
\ppim categorizes the peripheral registers based on their access patterns.
However, register mis-categorization could happen as acknowledged
by the \ppim authors.
For the firmware samples provided by \ppim~\cite{p2imgit},
register mis-categorization merely slowed down the fuzzing process and affected coverage improvement.
For others, we found that mis-categorization actually
severely influenced the usability of \ppim.
That being said, we did observe failed emulations with \ppim.
For example, in the \texttt{RF\_Door\_Lock} firmware,
\ppim mistakenly categorized the RCC register as control register which
actually should be a combination of control and status register.
As a result,
\ppim always returned the last written value to this register which
cannot satisfy the firmware expectation and 
eventually hung the execution.
In addition,
\ppim groups registers based on spatial adjacency.
Registers within \texttt{0x200} bytes are considered to belong to the same peripheral.
This assumption is not applicable for complex peripherals like USB, CAN and Radio Controller,
which have large or separated range.
This also leads to register mis-categorization.
The sample \texttt{Thermostat}, \texttt{LiteOS\_IoT} and \texttt{Zepyhr\_SocketCan} also stalled
during emulation due to register mis-categorization.

\paragraph{IA~--~Invalid Assumption about Registers.}
\ppim models a special kind of register which combines the functionality
of the control register and the status register.
It assumes that the control bits and status bits do not overlap. 
However, we found this assumption does not always hold.
For example, on the STM32F103RE chip,
the first bit of a register in the \textit{ADC} peripheral
is used as both the control bit and status bit.
The \texttt{3DPrinter} firmware sets this bit as one
and then waits for it to become zero.
Since \ppim recognized this bit as control bit,
it always returned one, making
the firmware stalled.
The same occurred to the sample used in \texttt{XML\_Parser}.




\paragraph{LE~--~Limited Exploration.}
\ppim cannot find appropriate values for status registers based on existing heuristics.
Therefore, it proposes \emph{explorative execution}. 
Specifically, it pauses and snapshots the
execution at register reading points. 
Then, \ppim spawns a worker thread for each candidate value.
The worker thread runs
with the assigned candidate and terminates when
it is about to return to the next level callee. 
Finally, the best value which
does not crash or stall the execution is picked. 
The problem with explorative execution
is that it is impossible to try all the candidates in the search space, because there could be as many as $2^{32}$ candidates for a peripheral register in a 32-bit MCU.
\ppim simply narrows down the search space by only investigating candidates
with a single bit set, meaning that only 32 plus 1 candidates are checked.
However, based on our experiments, multi-bit status registers are quite common, especially in complex peripherals like CAN and USB.
For example, the two samples used in HALucinator use
the \textit{SYSCTRL} peripheral to control
device oscillators and clock sources.
When the firmware enables the \texttt{DFLL48M} (i.e., Digital Frequency Locked Loop) feature,
a multi-bit status register (at \texttt{0x40000080C}) is in use.
\ppim cannot find the expected values, so the emulation was stalled.
\vspace*{-3mm}
\section{Limitations}
\vspace*{-2mm}
\label{sec:dma}

\label{sec:limitation}


Leveraging symbolic execution, \sys can achieve
dependency-aware peripheral access handling and constraint-satisfaction-based response finding. This enables \sys to use less heuristics but achieve better
accuracy compared with other works. However, when heuristics fail, there are
still some corner cases and human efforts are needed.



First, the proposed invalidity checking might not cover all invalid states.
Ideally, a proper implementation should check the error code
immediately after peripheral operations and
handle the exception, \eg~by letting the firmware enter an infinite loop. 
However, if the firmware continues normal execution, 
\sys cannot distinguish which branch target is better and  
have to randomly selected one.
We show such an example in Listing~\ref{lst:i2ctestsample}.
In this example, fuzzing test-cases cannot be fed to the emulator 
via the data register of the I2C peripheral.
As a result, bugs caused by inputs from the I2C peripheral cannot be discovered.
To deal with this kind of false negatives,
the analyst needs to provide user-defined program points 
that \sys should avoid reaching. 
Note that analysts can examine the log information generated by \sys
to quickly find out this information.


Second, \sys may mistakenly treat a legitimate long loop
as invalid state.
For example, we observed standard string and memory functions such as \texttt{memset} 
was sometimes detected as an invalid long loop.
This is because the contexts of these function invocations
involve symbolic values, which activate the invalid state
detector plugin.
Such false positives can be easily detected since all valid paths inevitably pass this point during firmware execution.
Therefore, if we found many paths fall in invalid states at the same point due to same reason, we can exclude these points from invalid states detection. 

Third, we rely on the characteristics described in Section~\ref{sec:fuzzing}
to identify data registers. However, we did observe rare cases when a true
data register does not exhibit these characteristics.
If a data register is mis-categorized as a \tone or \ttwo type,
\sys would only respond to it with a few fixed values and
the fuzzer cannot reach paths that
depend on the input from the data register.
In our evaluation, this rarely occurs.
As shown in Table~\ref{tab:configuration}, we missed only three out of 43 data registers.
Note that the 43 data registers were identified by reviewing the chip manual
and therefore can serve as the ground-truth. 
If a false negative is discovered, 
we allow analysts to directly add additional data registers via 
the configuration file.

In addition, \sys detects infinite/long loops only if the processor context
contains one or more symbols. However, it might happen that the counter of a
long loop is a concrete value but is dependent upon a symbol outside 
the loop. \sys would
miss the detection of this long loop because all the registers in the loop are concrete values.
Fortunately, we did not observe any such cases in our experiments. Considering the
diversity and complexity of real-world firmware, we acknowledge 
this limitation.

\vspace*{-2mm}
\section{Related Work}
\vspace*{-2mm}


To enable executing MCU firmware in an emulated
environment, most of the previous works~\cite{muench2018avatar2, koscher2015surrogates, kammerstetter2014prospect} follow a hybrid emulation approach, which forward the peripheral
access requests to the real hardware.
However, this approach suffers from poor performance.
M.Kammerstetter~\etal~\cite{kammerstetter2016embedded}
propose utilizing a cache for peripheral device communication
to improve the performance.
However, hardware-in-the-loop approaches are
not scalable for testing large-scale firmware images.
Instead of
fetching data from real devices, our approach infers proper
inputs with symbolic execution.

Recently, several research efforts~\cite{gustafson2019toward,fengp2020p2im, clements2020halucinator,cao2020device} have been focused on
firmware emulation
without hardware dependence.
Similar to \sys,
Laelaps~\cite{cao2020device} also uses the symbolic execution to infer appropriate responses to unknown peripheral accesses.
However, Laelaps only stays in symbolic execution mode for a short period (less than six basic blocks based on the paper)
before the path explosion problem begins to influence its performance.
Therefore, a peripheral input, after six basic blocks,
has to be concertized and cannot be involved in constraint solving.
In other words,
Laelaps can only find the ``best'' short-term path,
which may not be a valid path in the long run.
In addition, the architecture of Laelaps does not support caching.
Every access to peripherals traps the system into the symbolic execution engine,
leading to unacceptable performance overhead.
For example, in fuzzing the synthesized vulnerable firmware,
Laelaps executed less than 1,000 test-cases in an hour~\cite{cao2020device}.
The low performance makes it very inefficient in fuzzing, which relies on \emph{executions
per second}.

PRETENDER~\cite{gustafson2019toward} observes interactions between the
hardware and firmware, and uses machine learning and pattern recognition
to create models of peripherals.
Thus, it needs real devices to collect
the interactions between the original
hardware and firmware, and then learns the behavior.
This approach is less scalable if the firmware was written
for unpopular MCUs. Moreover,
the analyzed firmware cannot activate more peripheral features apart from
those already learned on real devices.

\ppim~\cite{fengp2020p2im} generates responses to peripheral accesses based on the categorization 
information of the peripheral.
It observes the access pattern of peripherals and relies
expert-provided heuristics to categorize each peripheral register.
We discuss how mis-categorization influences the accuracy of \ppim in
handing complex peripherals like USB, CAN and Radio in Section~\ref{sec:ppimfailreasons}.
Moreover, it cannot generate responses for many kinds of registers, in particular status registers.
This is because \ppim uses a concrete exploration algorithm to guess valid readings of registers,
while the huge search space makes it impractical.
For example,
if the firmware waits for a status register to have multiple bits set, \ppim can never find the expected value as discussed in Appendix~\ref{sec:ppimfailreasons}.

HALucinator~\cite{clements2020halucinator}
avoids peripheral emulation by replacing the high-level 
\textit{hardware abstraction layer} (HAL)
functions with a host implementation.
In this sense, it does not really model peripherals.
Therefore, comparing \hal with \sys, \ppim or Laelaps
is not perfectly fair.
Since HAL functions are replaced by 
host functions, it does not need to consider low-level
implementation, such as DMA.
However, since low-level drivers are skipped
for emulation, bugs resting there can never be exposed.
Also, building a database that matches all HAL libraries
needs the HAL source code
from all the major MCU vendors.
As a result, the wide adaptation of \hal
demands collaboration from industry.
SoCs with proprietary SDKs (e.g., Samsung SmartThings~\cite{SmartThings} and Philips~\cite{Philips}) cannot be supported by \hal.
Given the clear advantages and disadvantages of
\hal and \sys/\ppim/Laelaps, we argue that a combination
could generate a state-of-the-art tool for analyzing MCU firmware.
We can first use \hal to match any HAL functions and hook them
with host implementations.
During run-time, if any unknown peripheral is accessed,
\sys, \ppim or Laelaps can kick in and emulate the rest.

Apart from the emulation capability itself,
a distinct advantage of \sys to related work is the tight integration with S2E,
a platform for software analysis.
Therefore, there are many excellent plugins which 
are readily available.
Also, analysts can develop new plugins for \sys so
that other dynamic analysis mechanisms can be integrated.

\vspace*{-4mm} 
\section{Conclusions}
\vspace*{-2mm} 
This paper presents \sys, a new tool to 
emulate firmware execution, for the purpose of finding bugs in task code of firmware,
with a focus on those caused by malformed inputs from I/O interfaces. 
It automatically 
finds appropriate responses for accesses to unknown peripherals,
allowing for executing MCU firmware in an emulated environment
without requiring real hardware.
Our algorithm leverages symbolic execution to find new
paths and uses invalidity checking to make sure that
the firmware execution does not enter an invalid state.
At the same time, \sys learns the appropriate values for peripheral access and store them into a knowledge base.
After the knowledge extraction phase, with the returned knowledge base,
\sys efficiently responds to peripheral reading operations for dynamic analysis.
We have implemented our idea on top of S2E and developed
a fuzzing plugin. Evaluation results show that \sys
is capable of emulating real-world firmware and finding new bugs.








\vspace*{-2mm}
\section*{Acknowledgments}
\vspace*{-2.5mm}
We would like to thank our shepherd William Enck
and the anonymous reviewers for their helpful feedback.
We thank Bo Feng for providing us with the firmware samples used in \ppim~\cite{fengp2020p2im} and kind guidance on configuring
\ppim.
We also thank Vitaly Chipounov for his help on adding ARM support to S2E.
Wei Zhou and Yuqing Zhang were support by National Natural Science Foundation of China (U1836210) and CSC scholarship.
Le Guan was supported in part by JFSG from the University of Georgia Research Foundation, Inc.
Peng Liu was supported by ARO W911NF-13-1-0421 (MURI), NSF CNS-1814679, and NSF CNS-2019340.

{\footnotesize
\bibliographystyle{plain}
\bibliography{main}

\begin{thebibliography}{10}

\bibitem{halgit}
{HALucinator firmware samples}.
\newblock \url{https://github.com/ucsb-seclab/hal-fuzz/tree/master/tests}.

\bibitem{p2imrepo}
{P2IM real-world firmware samples}.
\newblock \url{https://github.com/RiS3-Lab/p2im-real_firmware}.

\bibitem{p2imgit}
{P2IM unit test samples}.
\newblock
  \url{https://github.com/RiS3-Lab/p2im-unit_tests/tree/30e6aec9f5c44f11b8072bf597eb80729dad417d}.

\bibitem{pretednergit}
{Pretender firmware samples}.
\newblock
  \url{https://github.com/ucsb-seclab/pretender/tree/master/test_programs/max32600}.

\bibitem{DSPIdriver}
{Bug Report: Critical memory leak in DSPI Master Peripheral Driver in
  combination with FreeRTOS}.
\newblock
  \url{https://community.nxp.com/t5/Kinetis-Software-Development-Kit/Bug-Report-Critical-memory-leak-in-DSPI-Master-Peripheral-Driver/m-p/374518},
  2020.

\bibitem{IoTClub}
{LiteOS Partner Development Kits}.
\newblock \url{https://github.com/LiteOS/LiteOS_Partner_Development_Kits},
  2020.

\bibitem{utasker}
{$\mu$Tasker}.
\newblock \url{https://www.utasker.com/index.html}, 2020.

\bibitem{modbusdemo}
{$\mu$Tasker MODBUS Extension Module}.
\newblock \url{https://www.utasker.com/modbus.html}, 2020.

\bibitem{usbdemo}
{$\mu$Tasker USB Demo}.
\newblock \url{https://www.utasker.com/docs/uTasker/uTaskerV1.3_USB_Demo.PDF},
  2020.

\bibitem{s2esinglepath}
{S2E: A Platform for In-Vivo Analysis of Software Systems) Manufacturers for
  2020}.
\newblock \url{https://s2e.systems/}, 2020.

\bibitem{armsupport}
{S2E official Issue of ARM Support}.
\newblock \url{https://github.com/S2E/s2e-env/issues/268}, 2020.

\bibitem{SocketCan}
{Socket CAN Sample}.
\newblock
  \url{https://docs.zephyrproject.org/latest/samples/net/sockets/can/README.html},
  2020.

\bibitem{topmcu}
{Top 10 Microcontrollers (MCU) Manufacturers for 2020}.
\newblock
  \url{https://www.bisinfotech.com/top-10-microcontrollers-mcu-manufacturers-2020/},
  2020.

\bibitem{Zephyr}
{Zephyr}.
\newblock \url{https://www.zephyrproject.org/}, 2020.

\bibitem{cadar2008klee}
Cristian Cadar, Daniel Dunbar, Dawson~R Engler, et~al.
\newblock Klee: Unassisted and automatic generation of high-coverage tests for
  complex systems programs.
\newblock In {\em OSDI}, volume~8, pages 209--224, 2008.

\bibitem{cao2020device}
Chen Cao, Le~Guan, Jiang Ming, and Peng Liu.
\newblock Device-agnostic firmware execution is possible: A concolic execution
  approach for peripheral emulation.
\newblock In {\em Annual Computer Security Applications Conference}, pages
  746--759, 2020.

\bibitem{chen2016towards}
Daming~D Chen, Maverick Woo, David Brumley, and Manuel Egele.
\newblock Towards automated dynamic analysis for linux-based embedded firmware.
\newblock In {\em NDSS}, volume~16, pages 1--16, 2016.

\bibitem{chipounov2011s2e}
Vitaly Chipounov, Volodymyr Kuznetsov, and George Candea.
\newblock S2e: A platform for in-vivo multi-path analysis of software systems.
\newblock {\em ACM Sigplan Notices}, 46(3):265--278, 2011.

\bibitem{clements2020halucinator}
Abraham~A Clements, Eric Gustafson, Tobias Scharnowski, Paul Grosen, David
  Fritz, Christopher Kruegel, Giovanni Vigna, Saurabh Bagchi, and Mathias
  Payer.
\newblock Halucinator: Firmware re-hosting through abstraction layer emulation.
\newblock In {\em 29th USENIX Security Symposium}, pages 1--18, 2020.

\bibitem{costin2016automated}
Andrei Costin, Apostolis Zarras, and Aur{'e}lien Francillon.
\newblock Automated dynamic firmware analysis at scale: a case study on
  embedded web interfaces.
\newblock In {\em Proceedings of the 11th ACM on Asia Conference on Computer
  and Communications Security}, pages 437--448, 2016.

\bibitem{fengp2020p2im}
Bo~Feng, Alejandro Mera, and Long Lu.
\newblock P2im: Scalable and hardware-independent firmware testing via
  automatic peripheral interface modeling.
\newblock In {\em Proceedings of Usenix Security Symposium}, 2020.

\bibitem{garbelini2020sweyntooth}
Matheus~E Garbelini, Chundong Wang, Sudipta Chattopadhyay, Sun Sumei, and
  Ernest Kurniawan.
\newblock Sweyntooth: Unleashing mayhem over bluetooth low energy.
\newblock In {\em 2020 $\{$USENIX$\}$ Annual Technical Conference
  ($\{$USENIX$\}$ $\{$ATC$\}$ 20)}, pages 911--925, 2020.

\bibitem{opentracker}
GEOLINK.
\newblock {OpenTracker - 100\% Arduino compatible GPS/GLONASS vehicle tracker}.
\newblock \url{https://github.com/geolink/opentracker}, 2020.

\bibitem{godefroid2008automated}
Patrice Godefroid, Michael~Y Levin, David~A Molnar, et~al.
\newblock Automated whitebox fuzz testing.
\newblock In {\em NDSS}, volume~8, pages 151--166, 2008.

\bibitem{gustafson2019toward}
Eric Gustafson, Marius Muench, Chad Spensky, Nilo Redini, Aravind Machiry,
  Yanick Fratantonio, Davide Balzarotti, Aur{\'e}lien Francillon, Yung~Ryn
  Choe, Christophe Kruegel, et~al.
\newblock Toward the analysis of embedded firmware through automated
  re-hosting.
\newblock In {\em 22nd International Symposium on Research in Attacks,
  Intrusions and Defenses ($\{$RAID$\}$ 2019)}, pages 135--150, 2019.

\bibitem{liteOS}
HUAWEI.
\newblock {Huawei LiteOS}.
\newblock \url{https://www.huawei.com/minisite/liteos/cn/index.html}, 2020.

\bibitem{kammerstetter2016embedded}
Markus Kammerstetter, Daniel Burian, and Wolfgang Kastner.
\newblock Embedded security testing with peripheral device caching and runtime
  program state approximation.
\newblock In {\em 10th International Conference on Emerging Security
  Information, Systems and Technologies (SECUWARE)}, 2016.

\bibitem{kammerstetter2014prospect}
Markus Kammerstetter, Christian Platzer, and Wolfgang Kastner.
\newblock Prospect: peripheral proxying supported embedded code testing.
\newblock In {\em Proceedings of the 9th ACM symposium on Information, computer
  and communications security}, pages 329--340, 2014.

\bibitem{bugs_of_tcpip}
Ori Karliner.
\newblock {FreeRTOS TCP/IP Stack Vulnerabilities -- The Details}.
\newblock
  \url{https://blog.zimperium.com/freertos-tcpip-stack-vulnerabilities-details/},
  December 2018.

\bibitem{kim2020firmae}
Mingeun Kim, Dongkwan Kim, Eunsoo Kim, Suryeon Kim, Yeongjin Jang, and Yongdae
  Kim.
\newblock Firmae: Towards large-scale emulation of iot firmware for dynamic
  analysis.
\newblock In {\em Annual Computer Security Applications Conference}, pages
  733--745, 2020.

\bibitem{king1976symbolic}
James~C King.
\newblock Symbolic execution and program testing.
\newblock {\em Communications of the ACM}, 19(7):385--394, 1976.

\bibitem{koscher2015surrogates}
Karl Koscher, Tadayoshi Kohno, and David Molnar.
\newblock Surrogates: Enabling near-real-time dynamic analyses of embedded
  systems.
\newblock In {\em 9th USENIXWorkshop on Offensive Technologies (WOOT 15)},
  2015.

\bibitem{mera2020dice}
A.~Mera, B.~Feng, L.~Lu, and E.~Kirda.
\newblock Dice: Automatic emulation of dma input channels for dynamic firmware
  analysis.
\newblock In {\em 2021 2021 IEEE Symposium on Security and Privacy (SP)}, pages
  302--318, Los Alamitos, CA, USA, may 2021. IEEE Computer Society.

\bibitem{muench2018avatar2}
Marius Muench, Dario Nisi, Aur{\'e}lien Francillon, and Davide Balzarotti.
\newblock Avatar2: A multi-target orchestration platform.
\newblock In {\em Proc. Workshop Binary Anal. Res.(Colocated NDSS Symp.)},
  volume~18, pages 1--11, 2018.

\bibitem{muench2018you}
Marius Muench, Jan Stijohann, Frank Kargl, Aur{\'e}lien Francillon, and Davide
  Balzarotti.
\newblock What you corrupt is not what you crash: Challenges in fuzzing
  embedded devices.
\newblock In {\em NDSS}, 2018.

\bibitem{Philips}
Philips.
\newblock {Philips Hue}.
\newblock \url{https://www.philips-hue.com/en-us}, 2020.

\bibitem{ruge2020frankenstein}
Jan Ruge, Jiska Classen, Francesco Gringoli, and Matthias Hollick.
\newblock Frankenstein: Advanced wireless fuzzing to exploit new bluetooth
  escalation targets.
\newblock In {\em 29th $\{$USENIX$\}$ Security Symposium ($\{$USENIX$\}$
  Security 20)}, pages 19--36, 2020.

\bibitem{SmartThings}
Samsung.
\newblock {SmartThings Developer}.
\newblock \url{https://smartthings.developer.samsung.com/}, 2020.

\bibitem{asan}
Konstantin Serebryany, Derek Bruening, Alexander Potapenko, and Dmitriy Vyukov.
\newblock {AddressSanitizer: A Fast Address Sanity Checker}.
\newblock In {\em Proceedings of the 2012 USENIX Conference on Annual Technical
  Conference (ATC'12)}, 2012.

\bibitem{urgent11}
Ben Seri, Gregory Vishnepolsky, and Dor Zusman.
\newblock {Critical vulnerabilities to remotely compromise VxWorks, the most
  popular RTOS}.
\newblock Technical report, {ARMIS, INC.}, 2019.

\bibitem{shoshitaishvili2015firmalice}
Yan Shoshitaishvili, Ruoyu Wang, Christophe Hauser, Christopher Kruegel, and
  Giovanni Vigna.
\newblock Firmalice-automatic detection of authentication bypass
  vulnerabilities in binary firmware.
\newblock In {\em NDSS}, 2015.

\bibitem{talebi2018charm}
Seyed Mohammadjavad~Seyed Talebi, Hamid Tavakoli, Hang Zhang, Zheng Zhang,
  Ardalan~Amiri Sani, and Zhiyun Qian.
\newblock Charm: Facilitating dynamic analysis of device drivers of mobile
  systems.
\newblock In {\em 27th USENIX Security Symposium}, pages 291--307, 2018.

\bibitem{TP}
TP-LINK.
\newblock {KASA}.
\newblock \url{https://www.tp-link.com/us/kasa-smart/kasa.html}, 2020.

\bibitem{printer}
Erik van~der Zalm~et.al.
\newblock {Marlin Firmware}.
\newblock \url{https://marlinfw.org/}, 2020.

\bibitem{afl}
{Zalewski, Michal}.
\newblock {American Fuzzy Lop}.
\newblock \url{http://lcamtuf.coredump.cx/afl/}, 2010.

\end{thebibliography}
}

\appendix

\section{Number of Blocks Used for Coverage Calculation}
\label{sec:appendix4}
\begin{table}[t]
\caption{Number of Blocks Used for Coverage Calculation}
\centering
\begin{adjustbox}{max width=\columnwidth}
\begin{tabular}{c|l|r|r|r|}
\textbf{Refs}                         & \textbf{Firmware}          & \multicolumn{1}{c|}{\begin{tabular}[c]{@{}c@{}}\textbf{Reached TB \#}\\ \textbf{(uEmu)}\end{tabular}} & \multicolumn{1}{c|}{\begin{tabular}[c]{@{}c@{}}\textbf{Reached TB \#}\\ \textbf{(P2IM)}\end{tabular}} & \multicolumn{1}{l|}{\textbf{Total BB* \#}} \\ \hline
\multirow{10}{*}{\textbf{P2IM}}       & \textbf{CNC}               & 1621                                                                               & 1613                                                                               & 2424                            \\
                                      & \textbf{Console}           & 657                                                                                & 845                                                                                & 1830                            \\
                                      & \textbf{Drone}             & 1539                                                                               & 1282                                                                               & 1715                            \\
                                      & \textbf{Gateway}           & 2104                                                                               & 2176                                                                               & 3992                            \\
                                      & \textbf{Heat\_Press}        & 490                                                                                & 530                                                                                & 1622                            \\
                                      & \textbf{PLC}               & 505                                                                                & 524                                                                                & 1910                            \\
                                      & \textbf{Reflow\_Oven}       & 954                                                                                & 823                                                                                & 2354                            \\
                                      & \textbf{Robot}             & 1086                                                                               & 1177                                                                               & 2511                            \\
                                      & \textbf{Soldering\_Iron}    & 1709                                                                               & 1338                                                                               & 2756                            \\
                                      & \textbf{Steering\_Control}  & 529                                                                                & 471                                                                                & 1623                            \\ \hline
\multirow{2}{*}{\textbf{HALucinator}} & \textbf{6LoWPAN\_Sender}   & 2648                                                                               & N                                                                                  & 5482                            \\
                                      & \textbf{6LoWPAN\_Receiver} & 2596                                                                               & N                                                                                  & 5481                            \\ \hline
\multirow{2}{*}{\textbf{Predenter}}   & \textbf{RF\_Door\_Lock}    & 683                                                                                & N                                                                                  & 2803                            \\
                                      & \textbf{Thermostat}        & 1007                                                                               & N                                                                                  & 3952                            \\ \hline
\textbf{WYC}                          & \textbf{XML\_Parser}        & 2187                                                                               & N                                                                                  & 8311                            \\ \hline
\multirow{6}{*}{\textbf{uEmu}}        & \textbf{GPS\_Tracker}      & 781                                                                                & N                                                                                  & 3268                            \\
                                      & \textbf{LiteOS\_IoT}       & 1213                                                                               & N                                                                                  & 1946                            \\
                                      & \textbf{Zepyhr\_SocketCan} & 2281                                                                               & N                                                                                  & 4811                            \\
                                      & \textbf{3Dprinter}         & 1145                                                                               & N                                                                                  & 6047                            \\
                                      & \textbf{utasker\_MODBUS}   & 1885                                                                               & N                                                                                  & 3126                            \\
                                      & \textbf{utasker\_USB}      & 1239                                                                               & N                                                                                  & 2952                           
\end{tabular}
\end{adjustbox}
\label{tab:covnum}
\flushleft
\scriptsize{
*: We use IDA pro 7.0 to calculate the total basic block number. 
}
\end{table}

\begin{table}[t]
\caption{Number of entries for each cache type in the \kb}
\centering
\begin{adjustbox}{max width=\columnwidth}
\begin{tabular}{l|ccccc|ccccc}
                           & \multicolumn{5}{c}{\textbf{Regs. Read by Firmware}}                      & \multicolumn{5}{c}{\textbf{Conditional Regs. Read by Firmware}}            \\
\textbf{Firmware}          & \textbf{T0}  & \textbf{T1}  & \textbf{T2} & \textbf{T3} & \textbf{Total} & \textbf{T0} & \textbf{T1} & \textbf{T2} & \textbf{T3} & \textbf{Total} \\ \hline
\textbf{CNC}               & 37           & 8            & 0           & 3           & 48             & 3            & 8            & 0            & 3            & 14             \\
\textbf{Console}           & 12           & 11           & 0           & 0           & 23             & 2            & 11           & 0            & 0            & 13             \\
\textbf{Drone}             & 32           & 6            & 0           & 1           & 39             & 2            & 5            & 0            & 1            & 8              \\
\textbf{Gateway}           & 47           & 15           & 0           & 1           & 63             & 6            & 15           & 0            & 1            & 22             \\
\textbf{Heat\_Press}       & 5            & 14           & 0           & 1           & 20             & 0            & 14           & 0            & 1            & 15             \\
\textbf{PLC}               & 14           & 4            & 0           & 0           & 18             & 1            & 4            & 0            & 0            & 5              \\
\textbf{Reflow\_Oven}      & 27           & 8            & 0           & 0           & 35             & 4            & 8            & 0            & 0            & 12             \\
\textbf{Robot}             & 19           & 4            & 1           & 1           & 25             & 2            & 4            & 1            & 1            & 8              \\
\textbf{Soldering\_Iron}   & 38           & 14           & 1           & 1           & 54             & 11           & 14           & 1            & 1            & 27             \\
\textbf{Steering\_Control} & 6            & 17           & 0           & 0           & 23             & 0            & 16           & 0            & 0            & 16             \\
\textbf{6LoWPAN\_Sender}   & 18           & 29           & 0           & 1           & 48             & 3            & 29           & 0            & 1            & 33             \\
\textbf{6LoWPAN\_Receiver} & 18           & 29           & 0           & 1           & 48             & 3            & 29           & 0            & 1            & 33             \\
\textbf{RF\_Door\_Lock}    & 21           & 14           & 2           & 1           & 38             & 5            & 11           & 2            & 1            & 19             \\
\textbf{Thermostat}        & 19           & 18           & 2           & 1           & 40             & 5            & 18           & 2            & 1            & 26             \\
\textbf{XML\_Parser}       & 26           & 11           & 0           & 0           & 37             & 3            & 10           & 0            & 0            & 13             \\
\textbf{GPS\_Tracker}      & 11           & 16           & 0           & 1           & 28             & 1            & 14           & 0            & 1            & 16             \\
\textbf{LiteOS\_IoT}       & 36           & 9            & 2           & 0           & 47             & 2            & 8            & 2            & 0            & 12             \\
\textbf{Zepyhr\_SocketCan} & 23           & 12           & 0           & 0           & 35             & 1            & 11           & 0            & 0            & 12             \\
\textbf{3Dprinter}         & 26           & 16           & 0           & 1           & 43             & 1            & 16           & 0            & 1            & 18             \\
\textbf{\bm{$\mu$}tasker\_MODBUS}   & 43           & 18           & 1           & 0           & 62             & 5            & 11           & 1            & 0            & 17             \\
\textbf{\bm{$\mu$}tasker\_USB}      & 29           & 31           & 0           & 1           & 61             & 5            & 17           & 0            & 1            & 23             \\ \hline
                           & \textbf{517} & \textbf{304} & \textbf{9} & \textbf{15} & \textbf{835}   & \textbf{65}  & \textbf{273} & \textbf{9}  & \textbf{15}  & \textbf{362}  
\end{tabular}
\end{adjustbox}
\label{tab:kbentries}
\end{table}

\vspace*{-2mm}
\section{Summary of Cache Types in the \kb}

In Table~\ref{tab:kbentries}, for each sample, we summarize the number of entries for each cache type.


\vspace*{-4mm}
\section{Details of Real Firmware Samples}
\label{sec:appendix3}

\begin{table*}[t]
\caption{Details of real-world firmware samples}
\centering
\begin{adjustbox}{max width=\textwidth}
\begin{tabular}{l|l|l|l}
\textbf{Firmware}          & \textbf{MCU}   & \textbf{OS/Sys lib.} & \textbf{On-chip Peripherals Used by Firmware}                                            \\\hline
\textbf{CNC}               & STM32F429ZI    & Bare metal           & TIM2,TIM4,USART2,PWR,SYSCFG,EXTI,GPIOA,GPIOB,GPIOD,GPIOE,RCC,FLASH                       \\
\textbf{Console}           & MK64FN1M0VLL12 & RIOT                 & RTC,SIM,PORTA,PORTB,PORTE,WDOG,MCG,UART,SMC,GPIOB,GPIOE                                  \\
\textbf{Drone}             & STM32F103RB    & Bare metal           & TIM2,TIM3,TIM4,I2C1,GPIOA,GPIOB,GPIOC,TIM1,USART1,RCC,FLASH                              \\
\textbf{Gateway}           & STM32F103RB    & Arduino              & TIM1,TIM2,TIM3,TIM4,I2C1,AFIO,GPIOA,GPIOB,GPIOD,ADC1,GPIOC,RCC,FLASH,UART                \\
\textbf{Heat\_Press}       & SAM3X8E        & Arduino              & ADC,PMC,UART,CHIPID,EFC1,PIOA,PIOB,PIOC,PIOD,WDT                                         \\
\textbf{PLC}               & STM32F429ZI    & Arduino              & USART3,PWR,GPIOD,RCC,FLASH                                                               \\
\textbf{Reflow\_Oven}      & STM32F103RB    & Arduino              & USART2,AFIO,GPIOA,GPIOB,GPIOC,ADC1,RCC,FLASH                                             \\
\textbf{Robot}             & STM32F103RB    & Bare metal           & TIM2,I2C1,GPIOA,TIM1,USART1,RCC,FLASH                                                    \\
\textbf{Soldering\_Iron}   & STM32F103RB    & FreeRTOS             & TIM2,TIM3,IWDG,I2C1,AFIO,GPIOA,GPIOB,GPIOD,ADC1,TIM1,DMA1,RCC,FLASH                      \\
\textbf{Steering\_Control} & SAM3X8E        & Arduino              & TC,ADC,PMC,UART,CHIPID,EFC1,PIOA,PIOB,PIOC,PIOD,WDT                                      \\
\textbf{6LoWPAN\_Sender}   & SAM R21        & Contiki              & PORT,RTC,UART,I2C,TC3,TC4,   SPI,SYSCTRL,GCLK,PM,EIC,NVMCTRL,USB,MTB,RF233CTRL           \\
\textbf{6LoWPAN\_Receiver} & SAM R21        & Contiki              & PORT,RTC,UART,I2C,TC3,TC4,SPI,SYSCTRL,GCLK,PM,EIC,NVMCTRL,USB,MTB,RF233CTRL              \\
\textbf{RF\_Door\_Lock}    & MAX32600       & Mbed                 & GPIO,TIMER,UART,DAC0,DAC1,DAC2,DAC3,AFE,ICC,CLKMAN,PM,IOMAN                              \\
\textbf{Thermostat}        & MAX32600       & Mbed                 & GPIO,TIMER,UART,I2C,DAC0,DAC1,DAC2,DAC3,AFE,ICC,CLKMAN,PM,IOMAN                          \\
\textbf{XML\_Parser}       & STM32L152XE    & Mbed                 & TIM5,RTC,UART,PWR,PORTA,PORTC,RCC,FLASH                                                  \\
\textbf{GPS tracker}       & SAM3X/A        & Arduino              & UART0,UART1,ADC,EEFC0,WDT,PIO,CHIPID,SMC,USB,PM                                          \\
\textbf{LiteOS\_IoT}       & STM32L431      & LiteOS               & UART2,I2C,PWR,SYSCFG,EXTI,UART1,RCC,FLASH,GPIOA,GPIOB,GPIOC                              \\
\textbf{Zepyhr\_SocketCan} & STM32L432KC    & Zephyr               & TIM2,UART2,PWR,RCC,FLASH,GPIO,CAN                                                        \\
\textbf{3Dprinter}         & STM32F103RE    & Arduino              & TIM2,TIM3,TIM4,TIM5,TIM6,TIM7,PORTB,PORTA,PORTC,ADC1,ADC2,ADC3,UART,DMA                  \\
\textbf{\bm{$\mu$}tasker\_MODBUS}   & STM32F429ZIT6U & $\mu$tasker              & TIM2,TIM4,IWDG,UART2,UART3,PWR,UART1,SYSCFG,GPIOA,GPIOB,GPIOC,GPIOD,GPIOG,FLASH,ETHERNET \\
\textbf{\bm{$\mu$}tasker\_USB}      & STM32F429ZIT6U & $\mu$tasker              & TIM2,TIM4,IWDG,UART2,UART3,PWR,UART1,GPIOA,GPIOB,GPIOC,RCC,FLASH,USB                    
\end{tabular}
\end{adjustbox}
\label{tab:detail}

\end{table*}

\paragraph{GPS/GLONASS Vehicle Tracker.}
This is an open-source firmware for GPS/GLONASS vehicle tracker
provided by Geolink~\cite{opentracker}.
It includes many advanced features such as real-time tracking,  analog sensors, CAN bus, battery monitor, external commands, and many others.

\paragraph{Marlin 3D printer.}
Marlin~\cite{printer} is an open source firmware for the RepRap family of replicating rapid prototypers -- popularly known as ``3D printers''.

\paragraph{Lite OS IoT Demo.}
LightOS~\cite{liteOS} is the IoT OS released by the Huawei.
This is the IoT demo firmware for commercial testing multiple functions, including serial, sensor and NB-IoT communication module running on the IoTClub board~\cite{IoTClub}.

\paragraph{Socket CAN Test on Zephyr.}
Zephyr~\cite{Zephyr} is an RTOS built for IoT applications supported by Linux Foundation. 
This socket CAN sample~\cite{SocketCan} is a server/client application that sends and receives raw CAN frames using BSD socket API.

\paragraph{\bm{$\mu$}Tasker MODBUS and \bm{$\mu$}Tasker USB.}
$\mu$Tasker~\cite{utasker} is an embedded OS
with many$\mu$ready-made projects targeting a board class of embedded processors.
The MODBUS demo~\cite{modbusdemo} demonstrates the
use of the MODBUS extension with $\mu$Tasker over Ethernet (MOSBUS TCP).
The MODBUS protocol is widely used in PLC devices.
The USB demo~\cite{usbdemo} provides a menu-driven terminal interface via USB.
It can be commanded to operate as an \texttt{RS232} device.

\begin{table*}[t]
\caption{Configuration used in the experiment in Section~\ref{sec:experiment2}}
\centering
\begin{adjustbox}{max width=\textwidth}

\begin{tabular}{l|c|c|c|l|l}
\textbf{Firmware}          & \textbf{BB\#\_INV1} & \textbf{BB\#\_INV2} & \textbf{BB\#\_Term} & \textbf{User-defined Program Points} & \textbf{Automatically Identified Data Registers (Missed Ground Truth)} \\ \hline
\textbf{CNC}               & 30                  & 2,000                & 30,000                      & none                                                                                                                  & 0x40004404, 0x40020010, 0x40020C10                                                                     \\
\textbf{Console}           & 30                  & 2,000                & 30,000                      & none                                                                                                                  & 0x4006A007                                                                                 \\
\textbf{Drone}             & 30                  & 2,000                & 30,000                      & none                                                                                                                  & 0x40005410, 0x40013804                                                                                 \\
\textbf{Gateway}           & 30                  & 2,000                & 30,000                      & none                                                                                                                  & 0x40004404, 0x40005410                                                                                 \\
\textbf{Heat\_Press}       & 30                  & 2,000                & 30,000                      & none                                                                                                                  & 0x400E0818                                                                                             \\
\textbf{PLC}               & 30                  & 2,000                & 30,000                      & none                                                                                                                  & 0x40004804                                                                                             \\
\textbf{Reflow\_Oven}      & 30                  & 2,000                & 30,000                      & none                                                                                                                  & 0x40004404, 0x40010C08, 0x4001244C                                                                     \\
\textbf{Robot}             & 30                  & 2,000                & 30,000                      & none                                                                                                                  & 0x40005410                                                                                             \\
\textbf{Soldering\_Iron}   & 30                  & 2,000                & 30,000                      & none                                                                                                                  & 0x40005410, 0x40010808                                                                                 \\
\textbf{Steering\_Control} & 30                  & 2,000                & 30,000                      & none                                                                                                                  & 0x400E0818                                                                                             \\
\textbf{6LoWPAN\_Sender}   & 80                  & 2,000                & 30,000                      & i2c\_master\_wait\_for\_bus                                                                                                                  & 0x42001828, 0x42000828, (0x42000C28)                                                                               \\
\textbf{6LoWPAN\_Receiver} & 80                  & 2,000                & 30,000                      & i2c\_master\_wait\_for\_bus                                                                                                                 & 0x42001828, 0x42000828, (0x42000C28)                                                                               \\
\textbf{RF\_Door\_Lock}    & 30                  & 2,000                & 20,000                      & Mbed\_Die                                                                                                             & 0x40039020                                                                                             \\
\textbf{Thermostat}        & 30                  & 2,000                & 20,000                      & Mbed\_Die                                                                                                             & 0x40039020, 0x4010D800                                                                                 \\
\textbf{XML\_Parser}       & 30                  & 2,000                & 30,000                      & none                                                                                                                  & 0x40004C04                                                                                             \\
\textbf{GPS\_tracker}      & 30                  & 2,000                & 30,000                      & none                                                                                                                  & 0x40098018, 0x4009C018                                                                                 \\
\textbf{LiteOS\_IoT}       & 30                  & 2,000                & 30,000                      & none                                                                                                                  & 0x40004424, (0x48000010)                                                                               \\
\textbf{Zepyhr\_SocketCan} & 30                  & 2,000                & 30,000                      & none                                                                                                                  & 0x40004424, 0x400065B0-0x400065BC                                                                      \\
\textbf{3Dprinter}         & 30                  & 2,000                & 30,000                      & Fail\_Config                                                                                                          & 0x40013804                                                                                             \\
\textbf{\bm{$\mu$}tasker\_MODBUS}   & 30                  & 2,000                & 30,000                      & Error\_ConfigEthernet                                                                                                 & 0x40004804, 0x40004404, 0x40011004                                                                     \\
\textbf{\bm{$\mu$}tasker\_USB}      & 30                  & 2,000                & 30,000                      & none                                                                                                                  & 0x40004804, 0x40004404, 0x40011004, 0x50001000-0x50001FFC                                               
\end{tabular}

\end{adjustbox}
\label{tab:configuration}
\end{table*}

\section{Detailed Configuration of Each Tested Firmware Sample}
\label{sec:appendix2}
In Table~\ref{tab:configuration}, we list the detailed configuration of each sample.



\end{document}